\documentclass[
 reprint,
 superscriptaddress,
preprintnumbers,
 amsmath,amssymb,
 aps,
 prl,
floatfix,
]{revtex4-2}

\usepackage{graphicx}
\usepackage{dcolumn}
\usepackage{bm}

\usepackage[separate-uncertainty=false]{siunitx} 
\usepackage{dsfont}
\usepackage{amssymb}
\usepackage{amsmath}
\usepackage{amsfonts}
\usepackage{lipsum}
\usepackage{xurl}
\usepackage{textcomp}
\usepackage{amsmath}
\usepackage{comment}
\usepackage{diagbox}
\usepackage{xcolor}
\usepackage{cellspace}
\usepackage{makecell}
\setcellgapes{4pt}
\usepackage{longtable}
\usepackage{balance}
\DeclareMathAlphabet{\mathcale}{U}{eur}{m}{n}

\newcommand{\nuD}{\nu_0^\text{D}}
\newcommand{\deuterium}{D}
\newcommand{\hydrogen}{H}
\newcommand{\CPT}{\textit{CPT} }
\newcommand{\CPTe}{\textit{CPT}-even }
\newcommand{\CPTo}{\textit{CPT}-odd }

\begin{document}

\title{Demonstration of deuterium's enhanced sensitivity \\ to symmetry violations governed by the Standard-Model Extension}

\author{A.~Nanda}
\affiliation{Stefan Meyer Institute for Subatomic Physics, Dominikanerbastei 16, 1010 Vienna, Austria}
\affiliation{University of Vienna, Vienna Doctoral School in Physics, Universit\"atsring 1, Vienna, A-1010, Austria}

\author{D.~Comparat}
\affiliation{Universit\'e Paris-Saclay, CNRS, Laboratoire Aim\'e Cotton, 91405, Orsay, France}

\author{O.~Dulieu}
\affiliation{Universit\'e Paris-Saclay, CNRS, Laboratoire Aim\'e Cotton, 91405, Orsay, France}

\author{S.~Lahs}
\affiliation{Universit\'e Paris-Saclay, CNRS, Laboratoire Aim\'e Cotton, 91405, Orsay, France}
\affiliation{present address: Vienna Center for Quantum Science and Technology, Atominstitut, TU Wien, 1020 Vienna, Austria}

\author{C.~Malbrunot}
\affiliation{TRIUMF, 4004 Wesbrook Mall, Vancouver, BC V6T 2A3, Canada}
\affiliation{Physics Department, McGill University, Montréal, Québec H3A 2T8, Canada}
\affiliation{Physics and Astronomy, University of British Columbia, Vancouver BC, V6T 1Z1, Canada}

\author{L.~Nowak}
\affiliation{Stefan Meyer Institute for Subatomic Physics, Dominikanerbastei 16, 1010 Vienna, Austria}
\affiliation{University of Vienna, Vienna Doctoral School in Physics, Universit\"atsring 1, Vienna, A-1010, Austria}

\author{M.\! C.~Simon}
\email{corresponding author: martin.simon@oeaw.ac.at}
\affiliation{Stefan Meyer Institute for Subatomic Physics, Dominikanerbastei 16, 1010 Vienna, Austria}
\affiliation{present address: Marietta Blau Institute for Particle Physics, Dominikanerbastei 16, 1010 Vienna, Austria}

\author{E.~Widmann}
\affiliation{Stefan Meyer Institute for Subatomic Physics, Dominikanerbastei 16, 1010 Vienna, Austria}
\affiliation{present address: Marietta Blau Institute for Particle Physics, Dominikanerbastei 16, 1010 Vienna, Austria}


\begin{abstract}
We have performed hyperfine spectroscopy of two transitions in ground-state deuterium and searched for violations of \textit{CPT} and Lorentz symmetry that would manifest as sidereal variations of the observed transition frequencies.
Several nonrelativistic proton coefficients of the Standard-Model Extension framework have been addressed.
The spin-independent coefficients with momentum power $k$=2,4 are constrained for the first time.
Bounds on spin-dependent coefficients are improved by exploiting a sensitivity enhancement originating from the relative momenta of the nucleons in the deuteron.
The best previous constraints by hydrogen maser measurements are surpassed by 4 and 14 orders of magnitude for coefficients with $k$=2 and 4, respectively.
\end{abstract}

\keywords{Deuterium hyperfine structure, Rabi spectroscopy, Standard-Model Extension, symmetry violation, sidereal variations}

\maketitle


The additional neutron in deuterium ($^2$H;\deuterium ) with respect to hydrogen ($^1$H;\hydrogen ) leads, on the one hand, to minor changes of molecular bonds and to small isotope shifts~\cite{Pachucki1994}.
On the other hand, the masses of the two isotopes (relevant to metrology and neutrino physics~\cite{Rau2020}), their synthesis in the early universe (indicator for cosmological parameters~\cite{Mossa2020}), and their hyperfine structure~\cite{PhysRev.73.718} are quite distinct.
Consequently, there are compelling reasons for both comparative and complementary studies.
Those also extend to exotic versions of \hydrogen\ and \deuterium , where the electron is replaced by a muon~\cite{Pohl2010,Antognini2013,Pohl2016_muD,Gao2022_muH,HERNANDEZ2018377}, pion~\cite{GOTTA2004133,Strauch2010_piD}, kaon~\cite{PhysRevLett.78.3067,PhysRevLett.94.212302,Curceanu2019KaonicHDReview},
or antiproton~\cite{RICHARD1982349,WYCECH1985308,BAKER1988631,AUGSBURGER1999417,AUGSBURGER1999149}.
Such experiments gave rise to the proton and deuteron radii puzzles and probed quantum chromodynamics at low energies, thereby challenging and advancing the Standard Model (SM) of particle physics. \hydrogen\ masers were employed in high precision measurements to test beyond SM physics~\cite{Phillips2001,Humphrey2003} and, more recently, isotope shifts were recognized as a tool for placing bounds on new light-mass bosons~\cite{PhysRevA.108.052825}. 

Here, we exploit another significant difference between the two isotopes:
the proton momentum $\boldsymbol{p}$ from internal nuclear motion is by far higher in \deuterium\ ($\sim$\SI{0.1}{\giga \eV / c}) than in \hydrogen\ ($\sim$\SI{1}{\kilo \eV / c}).
As noted by Kosteleck\'y and Vargas a decade ago~\cite{Kostelecky2015} and elaborated for the present measurement lately~\cite{PhysRevD.109.055001}, this results in orders of magnitude higher sensitivity to specific violations of \textit{CPT} (combination of three discrete symmetries: Charge conjugation, Parity, and Time reversal) and Lorentz symmetry.
Accordingly, we report on Rabi-type measurements~\cite{PhysRev.53.318} of two hyperfine transitions in ground-state \deuterium\  and search for sidereal variations of the frequency signals in order to extract new and improved constraints within the Standard-Model Extension (SME) framework~\cite{PhysRevD.55.6760,PhysRevD.58.116002,PhysRevLett.82.2254}. \\

\begin{figure}[t]
    \centering
    \includegraphics[width=0.99\linewidth]{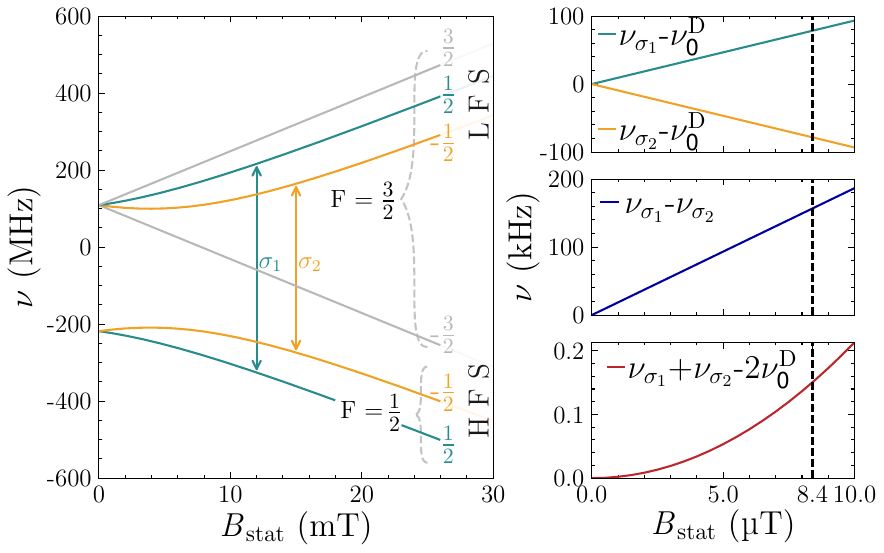}
    \caption{(left) Breit-Rabi diagram of deuterium showing the ground-state hyperfine structure and Zeeman shifts. The two $\sigma$ transitions ($\Delta M_F = 0$) are indicated with arrows and connect low-field-seeking (LFS) states of the quadruplet ($F=\frac{3}{2}; M_F=-\frac{3}{2},-\frac{1}{2},\frac{1}{2},\frac{3}{2}$) with high-field-seeking (HFS) states of the doublet ($F=\frac{1}{2}; M_F=-\frac{1}{2},\frac{1}{2}$). (right) Close-up views into the low magnetic field range relevant for this work ($B_\text{stat}$ = \SI{8.4}{\micro \tesla}, dashed line) show the Zeeman shifts for the $\sigma$ transitions as well as their combinations as difference and sum frequencies $\nu_\pm = \nu_{\sigma_1} \pm \nu_{\sigma_2}$.}
    \label{fig:D-HFS}
\end{figure}


The nuclear spin quantum number of \deuterium\ equals one.
Therefore, the ground-state hyperfine structure consists of a doublet and a quadruplet with corresponding total angular momentum quantum number $F$ of $\frac{1}{2}$ and $\frac{3}{2}$, respectively.
In the presence of an external magnetic field ($B_\text{stat}$), the degeneracy is lifted by different Zeeman shifts as described by the Breit-Rabi formula~\cite{PhysRev.38.2082.2,PhysRev.73.718} and shown in Fig.~\ref{fig:D-HFS}.
We investigate the two transitions with $\Delta M_F = 0$: $\sigma_1$ ($F,M_F: \frac{3}{2},\frac{1}{2} \rightarrow \frac{1}{2},\frac{1}{2}$) and $\sigma_2$ ($F,M_F: \frac{3}{2},-\frac{1}{2} \rightarrow \frac{1}{2},-\frac{1}{2}$), where $M_F$ is the magnetic quantum number.
The magnetic field dependence of these transition frequencies reads:
\begin{align}
\label{eq:sigma12}
\nu_{\sigma_1} \! = \! \nuD \sqrt{1 + \! ^2 \hspace{-0.05cm}/_3 \hspace{0.05cm} x + x^2}, \nonumber \\
\nu_{\sigma_2} \! = \! \nuD \sqrt{1 - \! ^2 \hspace{-0.05cm}/_3 \hspace{0.05cm} x + x^2},
\end{align}
with $\nuD$=\SI{327384352.5222(17)}{\hertz}~\cite{Wineland1972} the zero-field hyperfine splitting and $x=\mu^\text{D}_- B_\text{stat}/(h \nuD)$, where $h$ is Planck's constant and $\mu^\text{D}_- = - g_\text{e} \mu_\text{B} - g_\text{d} \mu_\text{N}$.
Here, $\mu_\text{B}$ ($\mu_\text{N}$) and $g_\text{e}$ ($g_\text{d}$) are the Bohr (nuclear) magneton and the electron (deuteron) $g$-factor, respectively.
We use 2022 CODATA values~\cite{Codata2022}, where the sign of $g_\text{e}$ is negative.
The field of $B_\text{stat}=\SI{8.4}{\micro \tesla}$ applied in this study corresponds to $x\simeq7\times 10^{-4}$. \\

The aforementioned SME framework generalizes the SM Lagrangian by adding operators violating Lorentz and \textit{CPT} symmetry.
Each operator of mass dimension $d$ is introduced with an associated coefficient of matching mass dimension ${4-d}$, and effects on observables, such as shifts of transition frequencies, can be calculated.
Thus, the SME enables quantitative comparability of various experiments contributing to a comprehensive and systematic search for symmetry violations~\cite{RevModPhys.83.11,SME-Tables-arXiv}.
Its initial minimal version~\cite{PhysRevLett.82.2254,PhysRevD.55.6760,PhysRevD.58.116002}
included Lorentz-violating operators with $d \leq 4$ and has been extended to incorporate arbitrary $d$ within the nonminimal SME~\cite{PhysRevD.80.015020,PhysRevD.88.096006,PhysRevD.99.056016}.
Combinations of coefficients for the same particle (i.e., flavor $\mathcale{w}$ like proton p, neutron n, or electron e), including those with different $d$, tend to appear together in calculations of observable effects and are conveniently collected as effective coefficients.
The example of relevance for the present work are the spherical nonrelativistic (NR) coefficients, where two types are distinguished:
\begin{align}
    \label{eq:coefficienttypes}
    \text{spin-dependent:} \ \mathcal{T}_{\mathcale{w}_{kjm}}^\text{ NR($qP$)} & =
    g_{\mathcale{w}_{kjm}}^\text{ NR($qP$)} - H_{\mathcale{w}_{kjm}}^\text{ NR($qP$)} , \nonumber \\
    \text{spin-independent:} \ \ \mathcal{V}_{\mathcale{w}_{kjm}}^\text{ NR} & =
    c_{\mathcale{w}_{kjm}}^\text{ NR}       - a_{\mathcale{w}_{kjm}}^\text{ NR} , 
\end{align}
with decompositions into \CPTo ($g,a$) and \CPTe ($H,c$) contributions.
Here, the index $k$ gives the momentum power, while $j$ and $m$ are the spherical tensor rank and component.
In brackets the spin weight $q$ and parity type $P$ of the operators are indicated, where parity types are denoted as $E$ and $B$-type for $(-1)^j$ and $(-1)^{j-1}$, respectively.
In this work the combinations ($0B$) and ($1B$) for spin-dependent and ($0E$) for spin-independent coefficients appear.

In principle, the present experiment on \deuterium\ tests e, n, and p coefficients.
However, for e coefficients, \deuterium\ gives no advantage over \hydrogen , as the electron's relative momenta are basically identical. 
Furthermore, stringent constraints for n coefficients exist from comagnetometry experiments, summarized in Tab.~VI of Ref.~\cite{PhysRevD.98.036003}.
Therefore, we concentrate on the subset of nonminimal NR p coefficients, where comagnetometry could not provide constraints yet, due to the lack of sufficiently elaborate nuclear structure modelling (see discussion in the End Matter).
The relationships between these coefficients and the energy shifts of the $\sigma$ transition are given by Eq.~(19) of Ref.~\cite{PhysRevD.109.055001}.
By forming their difference and sum ($\delta\nu_\pm = \delta\nu_{\sigma_1} \pm \delta\nu_{\sigma_2}$) the relations separate into spin-dependent and spin-independent components.
In natural units (speed of light and reduced Planck constant set to one: $c=\hbar=1$) we obtain:
\begin{align}
    \label{eq:nu-coeff-connect}
    2 \pi \ \delta \nu_- &
    = \frac{1}{6\sqrt{3\pi}}
    \sum_{_{\ q=0,1}^{k=0,2,4}} 
    (-1)^q \ \langle |\boldsymbol{p}|^k \rangle_{(qB)} \ \mathcal{T}_{p_{k10}}^{\text{ NR}(qB)} , \nonumber \\
    2 \pi \ \delta \nu_+ &
    = - \frac{1}{2\sqrt{5\pi}} \sum_{k=2,4}
    \langle |\boldsymbol{p}|^k \rangle_{(0E)} \ \mathcal{V}_{p_{k20}}^{\text{ NR}} ,
\end{align}
where $\langle |\boldsymbol{p}|^k \rangle_{(qP)}$ are the momentum expectation values of order $k$. For $k=2,4$, those are orders of magnitude larger in \deuterium\ than in \hydrogen~\cite{PhysRevD.109.055001}, and thus the source of the sensitivity enhancement exploited in this work.

In the inertial reference frame of the Sun, the SME coefficients are constant~\cite{RevModPhys.83.11,SME-Tables-arXiv}.
Their transformation to the coefficients acting in the Earth's frame is expressed by Eq.~(22) of Ref.~\cite{PhysRevD.109.055001}.
Concentrating on the terms relevant to this work we can reduce Eqs.~(25) of Ref.~\cite{PhysRevD.109.055001} to:
\begin{align}
    \label{eq:TVlabsuntransformReReduced}
    \mathcal{T}^{\text{ NR}(qB)}_{p_{k10}} = &
    - \sqrt{2} \ \sin{\vartheta} \ \Re \left( e^{i \omega_\oplus T_L} \ \mathcal{T}^{\text{ NR}(qB)\text{,Sun}}_{p_{k11}}
    \right) ,  \nonumber \\
    \mathcal{V}^{\textrm{ NR}}_{p_{k20}} = &
    - \sqrt{\frac{3}{2}} \ \sin{2\vartheta} \ \Re \left( e^{i \omega_\oplus T_L} \ \mathcal{V}^{\text{ NR,Sun}}_{p_{k21}}
    \right) \nonumber \\
    &
    + \sqrt{\frac{3}{2}} \ \sin^2{\vartheta} \ \Re \left( e^{i 2 \omega_\oplus T_L} \ \mathcal{V}^{\text{ NR,Sun}}_{p_{k22}} \
    \right) ,
\end{align}
where $\vartheta$ is the angle between the aligning static magnetic field $B_\text{stat}$ and the Earth's rotation axis, $\omega_{\oplus} \simeq 2 \pi/ (\SI{23}{\hour} \ \SI{56}{\min})$ is the sidereal frequency, and $T_L$ is the local sidereal time~\cite{PhysRevD.109.055001} defined such that the argument in the complex exponent can be written without constant phase offset.
The real ($\Re$) and imaginary ($\Im$) part of the coefficients in the Sun-centered frame would induce variations on the related coefficient in the Earth's frame following $\cos(m \omega_\oplus T_L)$ and $\sin(m \omega_\oplus T_L)$ functions, respectively. \\

\begin{figure}
    \centering
    \includegraphics[width=1\linewidth]{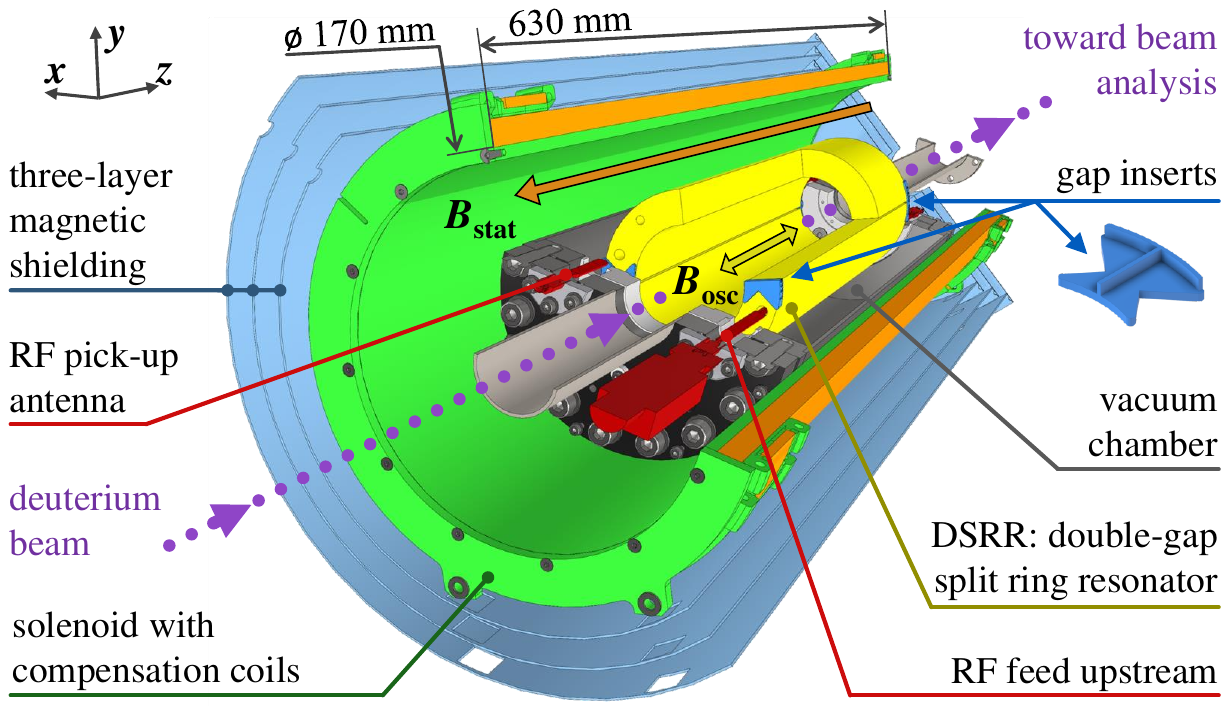}
    \caption{Sketch of the purpose-built cavity assembly. The static and oscillating magnetic fields ($\textbf{\textit{B}}_\text{stat} \parallel \textbf{\textit{B}}_\text{osc}$) needed to induce the $\sigma$ hyperfine transitions in \deuterium\ are indicated by the long arrow and short double-arrow, respectively.
    A solenoid with compensation coils in a three-layer magnetic shielding generates $\textbf{\textit{B}}_\text{stat}$.
    A double-gap split ring resonator in a vacuum enclosure provides $\textbf{\textit{B}}_\text{osc}$.
    Length and inner diameter dimensions are given for the main solenoid.
    By variation of the gap size through 3D-printed polymeric gap inserts the resonance frequency of the device was tuned to be close to $\nuD$. 
    }
    \label{fig:setup}
\end{figure}


A Rabi-type \hydrogen\ beamline has been constructed by the ASACUSA Collaboration to characterize equipment for antihydrogen hyperfine spectroscopy~\cite{Widmann200431,diermaier2017beam,MALBRUNOT2019110}.
It was optimized for \hydrogen\ velocities of $v_\text{H}\simeq\SI{1000}{\meter \second^{-1}}$, as this matched the anticipated antihydrogen beam properties. 
Upgrades to the initial apparatus, like combinations of ring apertures and sextupole magnets, which select narrow velocity ranges \mbox{($\Delta v_\text{H}\lesssim\SI{40}{\meter \second^{-1}}$)}, are succinctly described in~\cite{Nowak2024_PLB-HBeamSME,SuppMatA1}.

In the present work, we employ the existing source and detector.
The generation of a cold, modulated, and polarized atomic \deuterium\ beam merely requires supplying D$_2$ gas (instead of H$_2$), and for the detection, the quadrupole mass spectrometer selects a mass of \SI{2}{}~amu (instead of \SI{1}{}~amu).
The beam temperature and magnetic gradient force remain the same.
The corresponding scaling of velocities ($v_\text{D}\simeq\SI{700}{\meter \second^{-1}}$) and accelerations due to the mass difference between atomic \hydrogen\ and \deuterium\ results in trajectories which are basically identical for both species. \\

The cavity assembly, however, had to be replaced entirely.
As sketched in Fig.~\ref{fig:setup}, it consists of a vacuum chamber housing a radio-frequency (RF) resonator surrounded by a solenoid and a cylindrical three-layer magnetic shielding~\cite{SuppMatA2,SuppMatA3}.
The resonator employs a split ring geometry~\cite{Reynolds1991} to provide the oscillating magnetic field $B_\text{osc}$ around the frequency of $\nuD$ for stimulation of the hyperfine transitions.
Elongated versions of such resonators produce homogeneous axial oscillating magnetic fields within the inner cylindrical volume.
A distinctive feature of our device is a second gap, making it a double-gap split ring resonator (DSRR)~\cite{DSRR-1,DSRR-2}.
Gap size changes for frequency tuning are more flexible and avoid mechanical tension.
A signal generator connected to an in-phase power divider supplies two RF waves.
Those pass a weak, capacitive, over-coupled link and get fed into each shell of the DSRR from opposite ends through direct electrical contacts.
Coaxial to the feeding pin a pick-up antenna coupled to a spectrum analyzer monitors the RF field built up in the resonator.
The RF devices are locked to an external $\SI{10}{\mega \hertz}$ signal with $10^{-14}$ relative uncertainty, which is derived from the metrological network REFIMEVE~\cite{Kaur2022,Cantin_2021,RefimeveUrl}.

Inside the magnetic shielding layers, the solenoid with \SI{630}{\milli \meter} in length and \SI{170}{\milli \meter} inner diameter generates $B_\text{stat}$.
Compensation turns counteract the field drop at the entrance and exit of the solenoid.
Field mappings within the interaction volume  at $\sim$\SI{200}{\micro \tesla}~\cite{SuppMatA3} using fluxgate sensors confirmed values for the field inhomogeneity of $\varsigma_B / \overline{B} \simeq 5 \times 10^{-4}$, where $\varsigma_B$ refers to the standard deviation of the measured field values and overlines are used to denote averages from here on.
Currents are monitored as a proxy for the generated static field.
Fluxgate sensors provide direct magnetic field measurements inside and outside of the magnetic shielding.
Seven temperature sensors are mounted at critical positions to verify environmental stability. \\

The polarized D beam entering the interaction region consists of equal parts of the three LFS states (see Fig.~\ref{fig:D-HFS}).
Depending on the RF settings, a fraction of one of these states converts to a HFS state.
Prior to detection those atoms get removed from the beam by magnetic field gradients.
The signature for a resonant interaction is thus a drop in count rate, which at best can amount to one-third of the total beam rate.

Rabi oscillations were observed by setting the frequency to the center of a transition and scanning the RF power.
The first state population inversion corresponds to a $\pi$~pulse.
Supplied RF powers of \SI{-9.4}{\decibel}m and \SI{-10.1}{\decibel}m were found to be optimal for $\sigma_1$ and $\sigma_2$ transitions, respectively.
At these power settings, the frequency scans were performed.

\begin{figure}
    \centering
    \includegraphics[width=0.99\columnwidth]{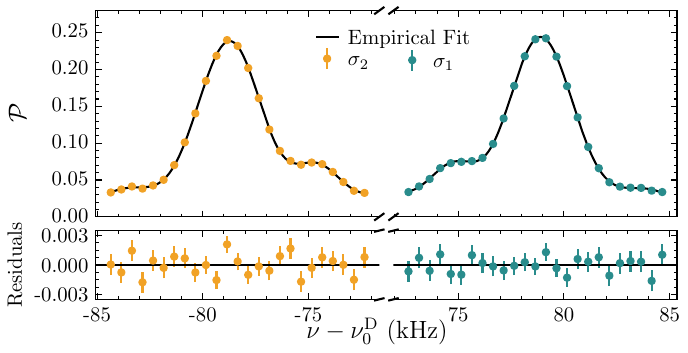}
    \caption{High-statistic transition probability spectra for the $\sigma_1$ (right) and $\sigma_2$ (left) transition by averaging all 107 resonance pairs of campaign \#2. The asymmetries originate from small inhomogeneities of $B_\text{stat}$. The two transitions feature a mirrored line shape due to their opposite Zeeman shifts.
    The line shows an empirical fit with 9 parameters, which is then used like a template to fit every resonance individually with only 3 free fit  parameters.
    Uncertainties are smaller than the dots and hence visualized as residuals in the lower panel.
    }
    \label{fig:lineshapes}  
\end{figure}

The acquisition of a \textit{resonance pair} consisted of point-wise interleaved scans of the $\sigma_1$ and $\sigma_2$ transition beam rates ($R_\sigma$), each sampled in a randomized sequence of 25 frequency points separated by \SI{500}{Hz}.
Prior to and following each $R_\sigma$, a reference measurement without RF interaction was performed ($R_\text{ref}$).
This enabled converting rates into probabilities through: $\mathcal{P}=1-R_\sigma/R_\text{ref}$~\cite{SuppMatB}. 

\begin{figure}
    \centering
    \includegraphics[width = 0.98\columnwidth]{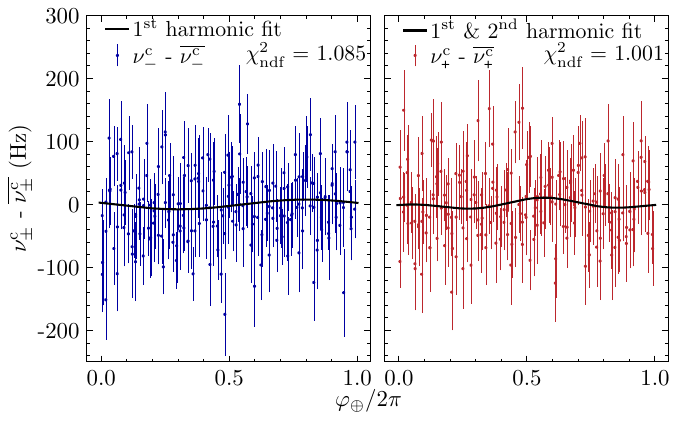}
    \caption{The difference (left) and sum (right) frequencies with offset corrections by subtracting averages ($\nu_\pm^\text{c}-\overline{\nu_\pm^\text{c}}$, for details see End Matter) are plotted against the sidereal phase. A fit (black line), according to Eq.~(\ref{eq:amplitudes}), extracts the values and statistical uncertainties of the amplitudes.
    }
    \label{fig:AmpFits} 
\end{figure}

Two campaigns of about one week duration each have been conducted in May (\#1) and August (\#2) 2023, accumulating 106 and 107 resonance pairs, respectively.
High-statistic probability spectra for both transitions are obtained by adding up all data of a campaign, as exemplified in Fig.~\ref{fig:lineshapes}.
The observed asymmetry is attributed to small inhomogeneities of $B_\text{stat}$ and is mirrored between the $\sigma_1$ and $\sigma_2$ transitions due to the opposite Zeeman shifts.
The high-statistic spectra are used to \emph{predetermine} 6 parameters of an empirical fit function with 9 parameters in total (see End Matter for details).
When applied to the individual resonance pairs only 3 fit parameters are free, namely a linear scaling of the function to the observed probabilities (two parameters) and the central frequency $\nu^\text{c}$. \\


Searches for sidereal variations are performed on the combined frequencies ${\nu^\text{c}_\pm = \nu^\text{c}_{\sigma_1} \pm \nu^\text{c}_{\sigma_2}}$ by plotting them against the mean sidereal phase ${\varphi_\oplus = \! \! \!\mod{(\omega_\oplus T_L,2\pi)}}$ of the data taking period for the corresponding resonance pair. 
Several amplitudes get extracted by fitting:
\begin{equation}
    \label{eq:amplitudes}
    \begin{split}
    \delta \nu_\pm (\varphi_\oplus)= A^\pm_0 + C^\pm_{1} \cos{\varphi_\oplus} + S^\pm_{1} \sin{\varphi_\oplus} + \\
    C^+_{2} \cos{2 \varphi_\oplus} + S^+_{2} \sin{2 \varphi_\oplus}\ .
    \end{split}
\end{equation}
Amplitudes for variations at the second harmonic are only relevant for $\mathcal{V}$ coefficients.
Therefore, those are only extracted from the sum frequencies $\nu^\text{c}_+$.
Figure~\ref{fig:AmpFits} displays these fits on the data of both campaigns.
The constant $A_0^\pm$ is removed by subtracting averages of the frequencies ($\overline{\nu_\pm^\text{c}}$), which are formed separately for three uninterrupted data taking periods (see End Matter for details on this \emph{offset correction}).

A relation between the SME coefficients in the Sun-centered frame and the amplitudes of sidereal variations of the hyperfine transitions measured in the Earth's frame is obtained by combining Eqs.~(\ref{eq:nu-coeff-connect}), (\ref{eq:TVlabsuntransformReReduced}), and (\ref{eq:amplitudes}).
Introducing a complex-valued amplitude $\mathcal{A}^{\pm}_m = C^{\pm}_m - i S^{\pm}_m$ and using natural units, we obtain:
\begin{align}
    \label{eq:amp2coeff-complex}
        2 \pi \ \mathcal{A}^-_{1} & = -\frac{\sin{\vartheta}}{3 \sqrt{6 \pi}} \sum_{_{\ q=0,1}^{k=0,2,4}} (-1)^q
                \langle |\boldsymbol{p}|^k \rangle_{(qB)} \  \mathcal{T}_{p_{k11}}^{\text{ NR}(qB)\text{,Sun}} ,  \nonumber \\
        2 \pi \ \mathcal{A}^+_{1} & = \frac{ \sin{2 \vartheta}}{2} \sqrt{\frac{3}{10 \pi}} \sum_{k=2,4}
                \langle |\boldsymbol{p}|^k \rangle_{(0E)} \  \mathcal{V}_{p_{k21}}^{\text{ NR}\text{,Sun}} ,  \nonumber \\
        2 \pi \ \mathcal{A}^+_{2} & = -\frac{\sin^2{\vartheta}}{2} \sqrt{\frac{3}{10 \pi}} \sum_{k=2,4}
                \langle |\boldsymbol{p}|^k \rangle_{(0E)} \  \mathcal{V}_{p_{k22}}^{\text{ NR}\text{,Sun}} . 
\end{align}
Note that the only complex-valued contributions on the right-hand sides are the SME coefficients.
The experiment was performed at the Laboratoire Aim\'e Cotton in Orsay near Paris at a latitude of \SI{48.7072\pm0.0002}{\degree}.
$B_\text{stat}$ is oriented \SI{6\pm2}{\degree} anticlockwise from due south when viewed from above, hence, $\vartheta$ = \SI{131.02\pm0.18}{\degree}~\footnote{The theoretical analysis toward the present experiment~\cite{PhysRevD.109.055001} assumes $\vartheta \simeq$\SI{49}{\degree} corresponding to the opposite magnetic field direction than applied in this work.}. \\

\begin{table}[t]
    \caption{Values with statistical, total, and systematic uncertainties (1~std.~dev.) for the amplitudes in units of \SI{}{\hertz} extracted by fits according to Eq.~(\ref{eq:amplitudes}), as shown in Fig.~\ref{fig:AmpFits}.}
    \label{tab:AmpResultsVar1}
        \renewcommand{\arraystretch}{1.3}
        \setlength{\tabcolsep}{2.8pt}
            \begin{tabular}{ccccccccccc}
                \hline \hline
                    ampl. (\SI{}{\hertz}) & & \multicolumn{3}{c}{$C^\pm_m=\Re(\mathcal{A}^\pm_m)$} & & \multicolumn{3}{c}{$S^\pm_m=-\Im(\mathcal{A}^\pm_m)$} & &common \\
                    $\pm$, harm. & & value & stat. & tot. & & value & stat. & tot. & & sys.  \\
                   \hline
                    $\mathcal{A}^-_\text{1}$  & & 2.5 \ &  5.0  &  6.2 & & -7.4 &  4.9 &  6.1 &  & 3.7  \\
                    $\mathcal{A}^+_\text{1}$ &  & -4.7 &  5.0  &  5.1 & &  -2.7 &  4.9 &  5.0 & & 1.2  \\
                    $\mathcal{A}^+_\text{2}$  &  & 3.8 &  4.9  &  4.9 & & 3.7 &  4.9 &  4.9 & & 0.3  \\
                    \hline \hline
            \end{tabular}
\end{table}

Systematic investigations evaluated various correlations between central frequencies $\nu^\text{c}$ extracted by the fits, RF power monitoring, temperatures, direct and indirect magnetic field measurements, as well as variations of the six predetermined parameters by three standard deviations of the empirical fit function and extraction of the linear independent amplitudes by fitting every term of Eq.~(\ref{eq:amplitudes}) separately to the data instead of simultaneously, as shown in Fig.~\ref{fig:AmpFits}.
However, all those tests showed negligible effects at the present level of statistics.
A search for other frequency components confirmed the absence of any significant signal in a wider range (see End Matter for details).
The impact of the aforementioned \emph{offset correction} was tested by rerunning the analysis without distinguishing uninterrupted data taking periods.
The absolute difference between the complex amplitudes $\mathcal{A}$ obtained with these two methods serves as a measure for systematic effects.
Amplitudes from the frequency difference ($\mathcal{A^-}$) appeared more susceptible to systematic effects than those of the sum frequency ($\mathcal{A^+}$), which is reasonable as the first order Zeeman shifts cancel for the latter. 
All values and uncertainties of amplitudes are summarized in Tab.~\ref{tab:AmpResultsVar1}, with the last column showing the potential systematic effects stemming from the offset correction, which is smaller than statistical uncertainties.

All constraints on SME coefficients derived from the amplitude limits are summarized in Tab.~\ref{tab:SMEconstraints}.
Details are exemplified and listed in~\cite{SuppMatD}.
The spin-dependent coefficients $\mathcal{T}_{p_{kjm}}^\text{ NR($qP$)}$ of momentum power ${k=0}$ are better constrained by \hydrogen\  maser measurements~\cite{Phillips2001,Humphrey2003,Kostelecky2015}, while those for $k=2$ and 4 are improved by 4 and 14 orders of magnitude.
This significant improvement can be entirely attributed to the use of \deuterium , taking advantage of the substantially larger proton momentum compared to \hydrogen , while it is still a similarly well-calculable system.  
The spin-independent coefficients $\mathcal{V}_{p_{kjm}}^\text{ NR}$ are constrained for the first time. \\

\begin{table}[t]
    \caption{Constraints on \textit{CPT}-even ($H,c$) and \textit{CPT}-odd ($g,a$) nonminimal NR SME proton coefficients in the Sun-centered frame as values with total uncertainties: (top section) For spin-dependent coefficients of momentum power $k=2,4$ the enhanced sensitivity of \deuterium\ enables significantly stronger constraints than achieved in \hydrogen\ maser measurements. (bottom section) Spin-independent coefficients are constrained for the first time through the sum frequency $\nu_+$.}
    \label{tab:SMEconstraints}
    \renewcommand{\arraystretch}{1.5}
        \begin{tabular}{lrrl}
            \hline\hline
                    spin-depend. coeff. & \multicolumn{1}{c}{ \ $\Re ( \mathcal{T}^\text{Sun} )$ } & \multicolumn{1}{c}{ \ $\Im ( \mathcal{T}^\text{Sun} )$ } & \multicolumn{1}{c}{units}  \\
                    \hline                
                    $H_{p_{211}}^\mathrm{  NR(0B)}$,  $-g_{p_{211}}^\mathrm{  NR(0B)}$   & $ 0.6 \pm 1.6$ & $ 1.9 \pm 1.6$ & \ \ \SI{e-20}{\giga \eV^{-1}} \\
                    $H_{p_{211}}^\mathrm{  NR(1B)}$,  $-g_{p_{211}}^\mathrm{  NR(1B)}$   & $ 1.5 \pm 3.7$ & $ 4.4 \pm 3.6$ & \ \ \SI{e-20}{\giga \eV^{-1}} \\
                    $H_{p_{411}}^\mathrm{  NR(0B)}$, $-g_{p_{411}}^\mathrm{ NR(0B)}$  & $ 1.8 \pm 4.6$ & $ 5.4 \pm 4.5$ & \ \ \SI{e-20}{\giga \eV^{-3}} \\
                    $H_{p_{411}}^\mathrm{  NR(1B)}$, $-g_{p_{411}}^\mathrm{  NR(1B)}$  & $ -0.5 \pm 1.1$ & $  -1.4 \pm 1.1$ & \ \ \SI{e-19}{\giga \eV^{-3}} \\
                    \hline
                    \hline
                    spin-independ. coeff. & \multicolumn{1}{c}{ \ $\Re ( \mathcal{V}^\text{Sun} )$ } & \multicolumn{1}{c}{ \ $\Im ( \mathcal{V}^\text{Sun} )$ } & \multicolumn{1}{c}{units}  \\
                    \hline
                    $c_{p_{221}}^\mathrm{  NR}$,         $-a_{p_{221}}^\mathrm{  NR}$      & $ 1.6 \pm 1.8$ & $ -0.9 \pm 1.7$ & \ \ \SI{e-20}{\giga \eV^{-1}} \\
                    $c_{p_{222}}^\mathrm{  NR}$,         $-a_{p_{222}}^\mathrm{  NR}$      & $ -2.3 \pm 3.0$ & $2.2 \pm 3.0$ & \ \ \SI{e-20}{\giga \eV^{-1}} \\
                    $c_{p_{421}}^\mathrm{  NR}$,      $-a_{p_{421}}^\mathrm{  NR}$   & $ -0.9 \pm 1.0$ & $ 0.5 \pm 1.0$ & \ \ \SI{e-19}{\giga \eV^{-3}} \\
                    $c_{p_{422}}^\mathrm{  NR}$,      $-a_{p_{422}}^\mathrm{  NR}$   & $1.3 \pm 1.7$ & $ -1.2 \pm 1.7$ & \ \ \SI{e-19}{\giga \eV^{-3}} \\
                    \hline \hline
        \end{tabular}
\end{table}


In summary, we have provided improved as well as new constraints on \textit{CPT} and Lorentz symmetry violations governed by nonrelativistic proton coefficients of the SME framework by searching for sidereal variations in combined frequencies of the $\sigma_1$ and $\sigma_2$ transition of the \deuterium\ ground-state hyperfine structure.
Despite inferior absolute precision in comparison to \hydrogen\  maser measurements, an improvement was possible due to a strong sensitivity enhancement mediated by the proton's momentum in the deuteron, thereby highlighting the potential of complementing \deuterium\ studies.

With further campaigns, the \deuterium\ experiment could provide additional constraints through a boost analysis as discussed in~\cite{PhysRevD.109.055001}.
Other opportunities are opened by measurements at various static magnetic field values and especially by reversing the field direction to address those SME coefficients, which do not lead to sidereal variations as showcased for \hydrogen\ just recently~\cite{Nowak2024_PLB-HBeamSME}.
Operation at lower velocities~\cite{Killian2023,killian2024} in the existing setup, introducing the Ramsey method~\cite{Ramsey1950,Ramsey1990experiments,Amit_CPT2019} in an improved setup, or a dedicated \deuterium\  maser~\cite{Wineland1972} could provide sensitivity enhancements of about one, two, or three orders of magnitude, respectively.

Finally, the community at the AD/ELENA facility of CERN (Antiproton Decelerator / Extra Low ENergy Antiproton ring) is presently evaluating future opportunities offered by antideuterons. Beyond the more general possibility of comparative antimatter studies to decouple limits on \CPTe and \CPTo coefficients, this system could give unique access to antineutron properties. \\


\textit{Acknowledgements---}We would like to express our gratitude to Arnaldo Vargas for laying the theoretical foundations this work is based on and for numerous enlightening discussions on the SME, to the staff of the SMI Advanced Instrumentation group for hardware support, to the technological platform of LAC (LAC Tech’), in particular Christophe Siour, for contributions to the transport of the experiment from CERN and its reinstallation at LAC, as well as to Simon Rheinfrank for valuable measurements on the static magnetic and RF fields.
We thank CERN's technical service groups for their continued support in general and specifically Manfred Wendt and the late Fritz Caspers for sharing their RF expertise, as well as Maxime Dumas, Luke Von Freeden, Mikko Karppinen, and the late Roberto Lopez for assisting our solenoid design. 
This project is supported by the European Unions Horizon 2020 research and innovation program under the Marie Sk\l odowska-Curie grant agreement No. 721559, the Austrian Science Fund FWF, Doctoral Program No. W1252-N27, and the TRIUMF start-up fund (Canada).
We acknowledge funding by the Austrian Academy of Sciences through the Investment Initiative.
This work was supported by program \textquotedblleft Investissements d’Avenir” launched by the French Government and implemented by ANR with the references ANR-21-ESRE-0029 (ESR/Equipex+ T-REFIMEVE and ANR-10-IDEX-0001-002 PSL (PSL).
Support from the CPER project COMB'IdF (Convention \#2022-IDF-P2 of Minist\`ere de l'Enseignement Sup\'erieur, de la Recherche et de l'Innovation, and of R\'egion Ile-de-France) is gratefully acknowledged.



\onecolumngrid
\subsection*{\large{End Matter}}
\twocolumngrid

\emph{SME sensitivity of other experiments---}High precision comagnetometry experiments employ even-odd noble gases.
Direct sensitivity to n related SME coefficients originates from the valence neutron, while sensitivity to p coefficients is suppressed.
So far only the Schmidt model~\cite{Schmidt1937} has been applied to the respective heavier nuclei in context of the SME framework.
In this model nucleons are paired to form states of zero total angular momentum, which results in a complete insensitivity of comagnetometry experiments to p coefficients~\cite{PhysRevD.98.036003}.
This model-dependent situation could change in the future if a residual sensitivity to SME p coefficients was quantified by more sophisticated nuclear structure modelling within the SME framework.

\begin{table*}
    \caption{Values with single standard deviations of six predetermined parameters of the empirical fit function~(\ref{eq:empirical_fit9}) obtained from fits to the high-statistic average spectra of each $\sigma$ transition and campaign, as shown in Figs.~\ref{fig:lineshapes}~and~\ref{fig:camp1HighStatFit}.
    In the final fits of resonance pairs, these six predetermined parameters are fixed to the mean values shown in the last column, leaving only three free parameter ($\nu^\text{c},\lambda,\kappa$).
    } 
    \label{tab:fitparameters1}
        \renewcommand{\arraystretch}{1.3}
        \setlength{\tabcolsep}{8pt}
            \begin{tabular}{l c c r r r r r}
                \hline \hline
                \multicolumn{3}{l}{parameter} &  \multicolumn{2}{c}{campaign \#1} & \multicolumn{2}{c}{campaign \#2} & \multicolumn{1}{c}{mean} \\
                name & symbol & units &
                \multicolumn{1}{c}{$\sigma_1$} & \multicolumn{1}{c}{$\sigma_2$} &
                \multicolumn{1}{c}{$\sigma_1$} & \multicolumn{1}{c}{$\sigma_2$} &  \\
                \hline
                Rabi asymmetry            & $\xi$   & 1             &
                0.0423(22) & -0.0402(23) & 0.0329(32) & -0.0323(32) & $\pm$0.0384  \\ 
                Rabi frequency            & $\Omega_\text{R}$ & \SI{}{\text{rad} \ \milli \second^{-1}} &
                11.07(22) & 11.05(22) & 9.62(43) & 9.87(38) & 10.77  \\ 
                interaction time          & $\tau_\text{int}$ & \SI{}{\milli \second} &
                0.2758(13) & 0.2753(14) & 0.2793(22) & 0.2808(22) & 0.2768  \\ 
                Lorentz probability ratio & $\rho$      & 1             &
                0.913(56) & 0.865(57) & 0.740(44) & 0.823(53) & 0.822  \\ 
                Lorentz frequency offset  & $\Omega_\text{L}$ & \SI{}{\text{rad} \ \milli \second^{-1}} &
                7.44(16) & -7.25(16) & 6.60(22) & -6.15(19) & $\pm$6.96  \\ 
                Lorentz width             & $\Gamma$      & \SI{}{\text{rad} \ \milli \second^{-1}} &
                51.6(25) & 49.8(27) & 37.0(19) & 41.3(26) & 43.5  \\ 
                \hline \hline
            \end{tabular}
\end{table*}

\emph{Empirical fit function---}In the present work, the hyperfine transition frequency is the quantity that is tested for variations during a sidereal day.
A line shape fit retrieves a value for the central transition frequency $\nu^\text{c}$ from the observed resonance spectra as a proxy for the true value $\nu^\text{c}_\text{true}$.
Understanding the physics behind the observed line shape in detail is useful but not critical.
For instance, a constant systematic offset $\Delta \nu_\text{sys}=\nu^\text{c}-\nu^\text{c}_\text{true}$ is of minor concern in a search for variations.
Consequently, it is sufficient to apply an empirical fit function.

Our Rabi-type spectroscopy measures the reduction in beam rate while a stimulating RF-field becomes resonant with the transition of interest.
The drop in rate depends on the applied RF frequency $\nu$ and is proportional to the transition probability $\mathcal{F}_\text{R}$ ideally given by:

\begin{equation}
\label{eq:RabiIdeal}
\mathcal{F}_\text{R}(\nu;\nu^\text{c},\Omega_\text{R},\tau_\text{int}) =
\frac{\sin^2 \left( \frac{1}{2} \Omega_\text{R} \tau_\text{int} \sqrt{1 + ( \Delta\Omega / \Omega_\text{R}  )^2} \right)}{1
+ ( \Delta\Omega / \Omega_\text{R} )^2} .
\end{equation}
The Rabi frequency $\Omega_\text{R}$ quantifies the constant interaction strength.
$\Delta\Omega=2\pi(\nu-\nu^\text{c})$ is the detune replacing the scan variable $\nu$ and the parameter $\nu^\text{c}$ on the right-hand side of the equation.
On resonance ($\Delta\Omega=0$), the first population inversion is achieved when the condition for a $\pi$~pulse is satisfied: $\Omega_\text{R} \tau_\text{int} = \pi$.
In the present experiment $\tau_\text{int}$ is given by the velocity of the \deuterium\ atoms and the length of the interaction apparatus and amounts to $\sim \SI{400}{\micro \second}$.
The Rabi frequency can be optimized to a $\pi$~pulse by adjusting the RF power $P_\text{RF}$ applied to the cavity due to the proportionalities $\Omega_\text{R} \propto B_\text{osc} \propto \sqrt{P_\text{RF}}$. 

Distortions of the line shape are caused by imperfections.
The following empirical modifications are made to Eq.~(\ref{eq:RabiIdeal}) to account for those.
The Rabi-type line shape is extended by an asymmetry parameter $\xi$, which enters by rewriting $\Delta\Omega / \Omega_\text{R} \rightarrow \Delta\Omega / ( \Omega_\text{R} + \xi \Delta\Omega) $.
Furthermore, a Lorentzian background is added: $\mathcal{F}_\text{L}=[ 1+(\Delta \Omega + \Omega_\text{L})^2 / \Gamma^2 ]^{-1}$, with a width $\Gamma$ and a frequency-offset $\Omega_\text{L}$.
$\mathcal{F}_\text{R}$ and $\mathcal{F}_\text{L}$ are combined with a weighing factor $\rho$ on the Lorentz contribution and then scaled to the directly measured probability ($\mathcal{P}_\text{meas}=1-R_\sigma/\overline{R}_\text{ref}$, compare main text) by a multiplier $\lambda$ and constant offset $\kappa$.
Hence, the resulting empirical probability fit-function has 9 parameters:
\begin{align}
& \mathcal{P}_\text{fit}(\nu;\nu^\text{c},\lambda,\kappa,\xi,\Omega_\text{R},\tau_\text{int},\rho,\Omega_\text{L},\Gamma) =  \nonumber \\ 
& \lambda \left[ \mathcal{F}_\text{R}(\Delta\Omega;\xi,\Omega_\text{R},\tau_\text{int}) + \rho \ \mathcal{F}_\text{L}(\Delta\Omega;\Omega_\text{L},\Gamma) \right]
+ \kappa \ .
 \label{eq:empirical_fit9}    
\end{align}
\begin{figure}[t]
    \centering
    \includegraphics[width=\linewidth]{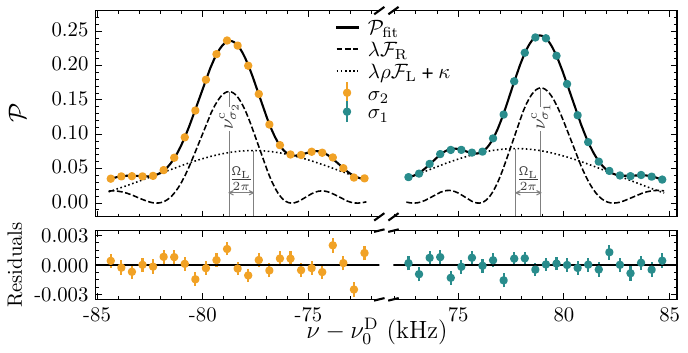}
    \caption{Campaign \#1 equivalent of Fig.~\ref{fig:lineshapes}. Dashed and dotted lines show Rabi- and Lorentz-type contributions to the empirical fit function given by Eq.~(\ref{eq:empirical_fit9}).}
    \label{fig:camp1HighStatFit}
\end{figure}
\begin{figure}[b]
    \centering
    \includegraphics[width=\linewidth]{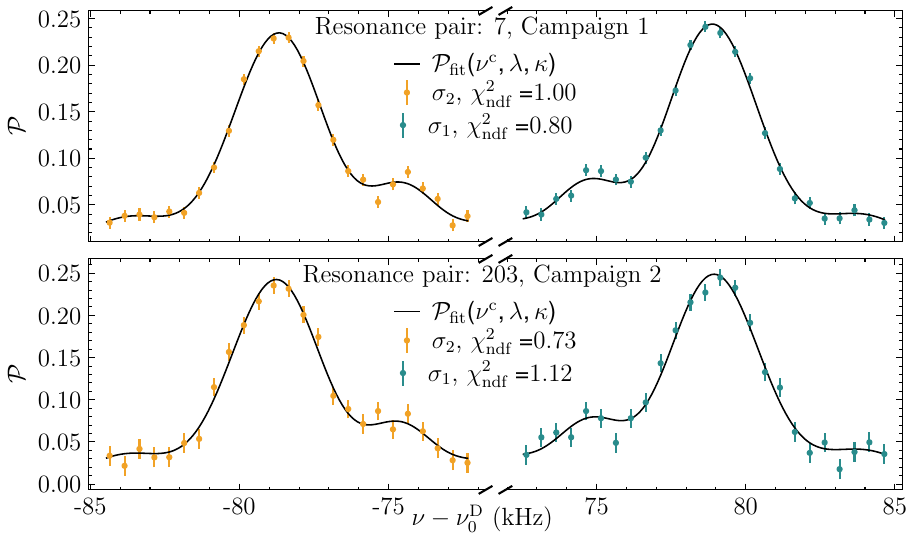}
    \caption{Examples of three-parameter fits to single resonance pairs from campaign \#1 and \#2 including typical values for reduced squared residuals ($\chi^2_\text{ndf}$).}
    \label{fig:threeparameterfitsexamples}
\end{figure}
In Fig.~\ref{fig:camp1HighStatFit}, the Rabi and Lorentz part of the fit function are illustrated. 
The parameters $\xi,\Omega_\text{R},\tau_\text{int},\rho,\Omega_\text{L}$, and $\Gamma$ are predetermined by fits to high-statistic spectra, which are obtained by averaging all spectra of each campaign for both $\sigma$ transitions.
The values are summarized in Tab.~\ref{tab:fitparameters1}.
A quantitative interpretation of Rabi-type line shape parameters is not meaningful, as the Lorentz-type background is introduced ad-hoc.
The parameters $\xi$ and $\Omega_\text{L}$ control asymmetries.
Therefore they are sign-opposite for the $\sigma_1$ and $\sigma_2$ transitions as their line shapes are mirrored due to sign-opposite Zeeman shifts.
Deviations between parameters for the two resonances of the same campaign are within uncertainties, while tensions exist between the campaigns.
Slight changes of static and RF field properties in combination with correlation between fit parameters can explain those.
The final analysis applies the empirical function with only three open fit parameters ($\nu^\text{c},\lambda,\kappa$) and the other six predetermined parameters fixed to the mean values tabulated in the last column of Tab.~\ref{tab:fitparameters1}.
Examples of this three-parameter fit are illustrated in Fig.~\ref{fig:threeparameterfitsexamples} for resonance pairs of each campaign.
The use of the parameter values of either campaign \#1 or \#2 produces the same results for the complex amplitudes within 5\% of the statistical uncertainty. \\

\emph{Systematics, offset corrections, data groups---}In spite of using the same current set values, the magnetic field was not identical in the two campaigns separated by three months.
In addition, a malfunction of the high-precision current supply after 82 out of 106 resonance pairs of campaign \#1 required a reset.
Therefore, the complex amplitudes $\mathcal{A}^-_{1}$, $\mathcal{A}^+_{1}$, and $\mathcal{A}^+_{2}$ are extracted after offset corrections on $\nu_\pm^\text{c}$ with three distinct frequency averages referred to as \#1a, \#1b, and \#2 in Tab.~\ref{tab:CampaignAverages}.
The complex amplitudes shift slightly when the offset correction is performed with a single joint frequency average instead (last column of Tab.~\ref{tab:CampaignAverages}).
The absolute difference $|\mathcal{A}^\text{indiv.}-\mathcal{A}^\text{joint}|$ yields the systematic errors listed in the last row of Tab.~\ref{tab:AmpResultsVar1}. \\

\begin{table}[h]
    \caption{
    Comparison of frequency averages $\overline{\nu_\pm^\text{c}}$ for groups of data (\#1a, \#1b, \#2). Top row states the respective number of acquired resonance pairs. Uncertainties are single standard deviations.
    }
    \label{tab:CampaignAverages}
        \renewcommand{\arraystretch}{1.3}
        \setlength{\tabcolsep}{2.5pt}
            \begin{tabular}{lcccc}
                \hline \hline
                 campaign: & \#1a & \#1b & \#2 & joint \\ 
                \hline
                acqu. res. pairs: & 82 & 24 & 107 & 213 \\
                \hline
                ($\overline{\nu_-^\text{c}} - \SI{157 000}{}$) \SI{}{\hertz}: &
                699$\pm$5 & 668$\pm$10 & 739$\pm$6 & 709.1$\pm$3.5 \\
                ($\overline{\nu_+^\text{c}} - 2\nuD$) \SI{}{\hertz}: &
                157$\pm$5 & 136$\pm$10 & 163$\pm$6 & 156.6$\pm$3.5 \\
                \hline \hline
            \end{tabular}
\end{table}

\emph{Lomb-Scargle periodograms---}
\begin{figure}[b]
    \centering
    \includegraphics[width=0.99\columnwidth]{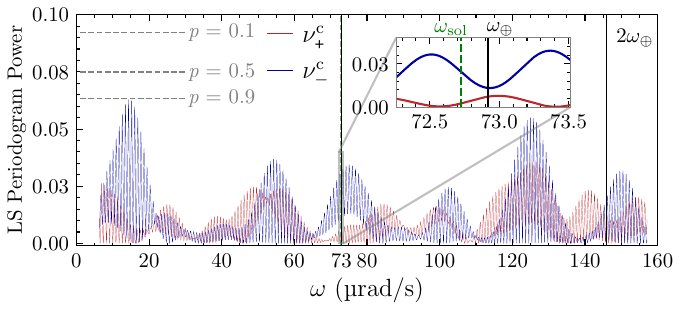}
    \caption{The graph shows standard normalized Lomb-Scargle periodograms for $\nu_\pm^\text{c}$ with horizontal dashed lines indicating $p$-values (details see text). The inset zooms into the frequency region of the sidereal and solar day.}
    \label{fig:LSP} 
\end{figure}
The presence of other frequency components in the time evolution of $\nu_\pm^\text{c}$ was tested by a Fourier transform method for unevenly sampled time series referred to as Lomb-Scargle periodograms~\cite{VanderPlas2018,VanderPlas2015,VanderPlas2012,Lomb1976,Scargle1982}.
Widely used by the astrophysics communities~\cite{astropy:2022,astropy:2018,astropy:2013}, the method has been recently applied in searches for sidereal variations to constrain Lorentz violations~\cite{Dreissen2022}.
The periodogram of the combined frequencies $\nu_\pm^\text{c}$ are shown in Fig.~\ref{fig:LSP}.
The frequency $\nu_-^\text{c}$ is more sensitive to changes of $B_\text{stat}$ that could be induced by environmental fluctuations, therefore the observation of higher power values in comparison to $\nu_+^\text{c}$ is reasonable.
Even the highest observed power levels can be explained with $\geq90\%$ probability as Gaussian noise when interpreted as $p$~values using the methods developed by Baluev~\cite{BaluevMethod}. 
The inset zooms onto the frequency region of interest.
The small fringes appear due to the three-month time separation between the two campaigns.
Day-night effects can thus be resolved from sidereal variations. \\


\onecolumngrid

\subsection*{\large{Supplemental Material}}

\renewcommand{\thefigure}{S.\arabic{figure}}
\setcounter{figure}{0}
\renewcommand{\thetable}{S.\Roman{table}}
\setcounter{table}{0}
\renewcommand{\thesubsection}{\Alph{subsection}} 
\renewcommand{\thesubsubsection}{\arabic{subsubsection}}
\setcounter{secnumdepth}{3}
\setcounter{tocdepth}{3}

\subsection{Experimental details}
\subsubsection{Beam path}

\begin{figure}[h] 
    \centering
    \includegraphics[width=\textwidth]{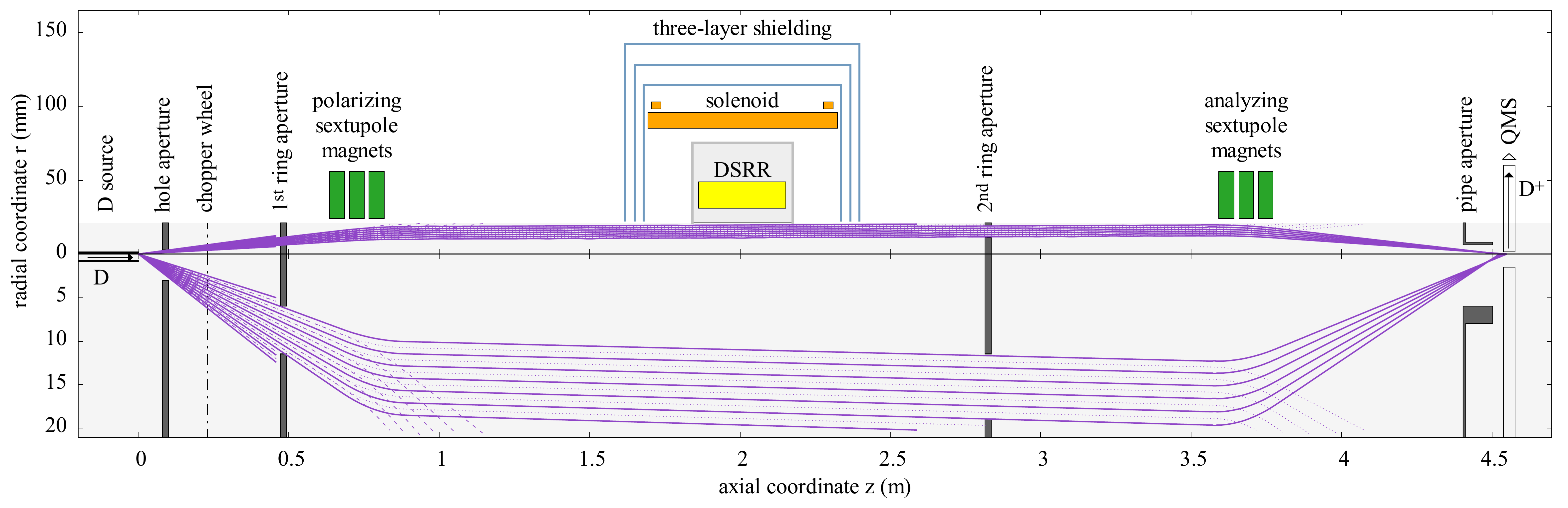}
    \caption{Sketch showing the arrangement of beam components to scale.  Purple lines are simulated trajectories for \deuterium\ of $v_\text{D}\simeq\SI{700}{\meter \second^{-1}}$. Note, that the $z$ direction is in \SI{}{\meter}, while $r$ is in \SI{}{\milli \meter} with different scales above and below the $r=0$ axis. Hence the section below $r=0$ is a mirrored zoom view onto the area filled by the beam of $r\lesssim\SI{20}{\milli \meter}$, which provides better visibility of the apertures as well as effects on trajectories by the polarizing and analyzing sextupole magnets. Trajectories of low/high-field-seeking states (LFS/HFS) are depicted by full/dashed lines. Dotted lines are for atoms, which transition from LFS to HFS when passing the hyperfine interaction region. See text for more details on the components.}
    \label{fig:beampath}
\end{figure}

Figure~\ref{fig:beampath} sketches components of the Rabi-type deuterium (\deuterium ) beam experiment to scale.
Simulated \deuterium\ trajectories depicted in purple start at the source on the left-hand and terminate at the quadrupole mass spectrometer (QMS) detector on the right-hand.
Molecular D$_2$ is dissociated in a plasma maintained by microwaves in an Evenson cavity~\cite{OpthosUrl}.
A PTFE tube of \SI{1.6}{\milli \meter} inner diameter, which is sandwiched between aluminium plates with good thermal contact to a cryocooler, guides and cools atomic \deuterium\ inside the first vacuum chamber.
We define the exit of this tube as the source point at the position $z=\SI{0}{\milli \meter}$.
A hole aperture with a diameter of \SI{6}{\milli \meter} at $z=\SI{0.09}{\meter}$ collimates the beam and
separates the cryogenic source chamber from a chamber housing the chopper at $z=\SI{0.25}{\meter}$.
A first ring aperture with inner and outer diameter of \SI{12}{\milli \meter} and \SI{23}{\milli \meter} at $z=\SI{0.50}{\meter}$ shapes an annular beam.
A set of three permanent sextupole magnets of \SI{50}{\milli \meter} in length each and centered at $z=\SI{0.78}{\meter}$ deflects high-field-seeking states (HFS: dashed trajectories) and bends low-field-seeking states (LFS: trajectories with full and dotted lines) roughly onto coaxial trajectories.
This produces a polarized beam, which enters the hyperfine interaction region described in more details in the subsequent paragraphs.
The velocity of the shown trajectories is $v_\text{D}\simeq\SI{700}{\meter \second^{-1}}$.
A second ring aperture with inner and outer diameters of \SI{23}{\milli \meter} and \SI{38}{\milli \meter} is located at $z=\SI{2.85}{\meter}$.
Another set of three permanent sextupole magnets, centered at $z=\SI{3.68}{\meter}$, analyses the beam by removing atoms which turned from LFS to HFS (dotted trajectories) by a hyperfine interaction.
Through a pipe aperture of \SI{15}{\milli \meter} inner diameter and \SI{100}{\milli \meter} in length centered at $z=\SI{4.48}{\meter}$ the beam reaches the detector chamber.
In a crossed-beams quadrupole mass spectrometer (QMS) at $z=\SI{4.57}{\meter}$, with a \SI{3}{\milli \meter} diameter opening, the atoms are ionized, mass-selected, and counted using a channeltron.

\subsubsection{Radio-frequency field generation}
\begin{figure}[h]
    \centering
    \includegraphics[width=0.7\columnwidth]{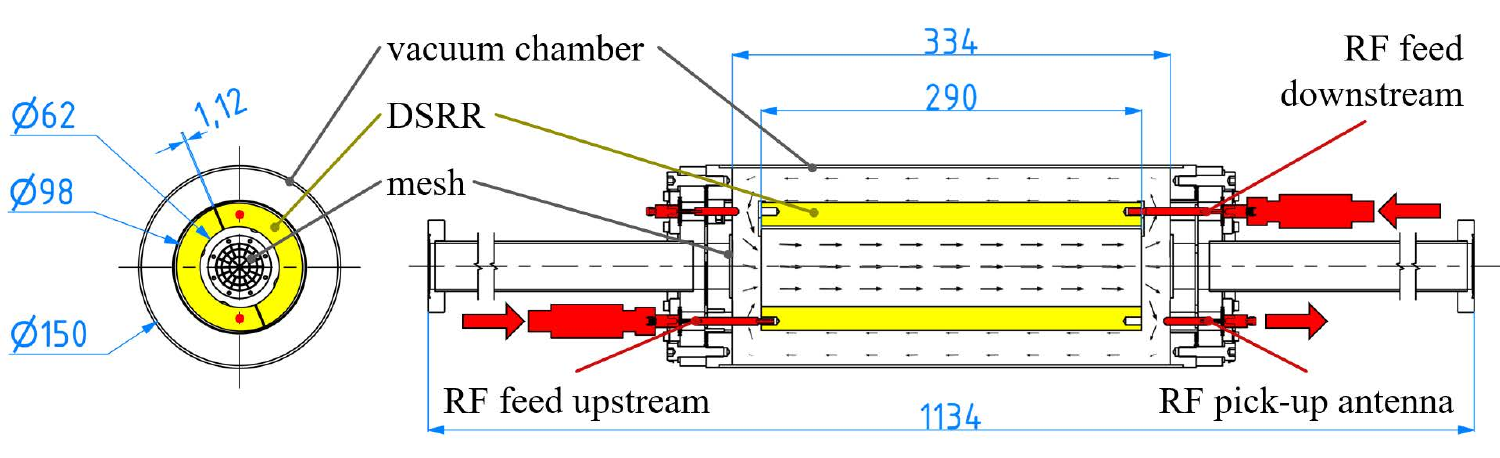}
    \caption{Dimensional cut views of the resonator (DSRR). $B_\text{osc}$ as obtained from Comsol~\cite{ComsolUrl} simulations is depicted by arrows in the longitudinal cut view. Radio frequency waves are fed over two pin connections and monitored via the downstream antenna. Dimensions are given in \SI{}{mm}.}
    \label{fig:DSRR-WireFrameComsol} 
\end{figure}

Recognizing the split ring as a single-turn coil with a capacitive gap reveals its equivalence to an LC resonator.
The resonance frequency can be tuned by changes of the gap size while the length of the device is of minor relevance allowing for extended homogeneous axial oscillating magnetic fields to be produced.
In addition the resonance wavelengths can be larger than the split ring dimensions making them an ideal choice for \deuterium\ hyperfine spectroscopy ($\lambda_0^\text{D} \sim \SI{91.6}{\cm}$)~\cite{Reynolds1991}.
Our DSRR made from aluminum has a length of \SI{290}{\milli \meter} as well as an inner and outer diameters of \SI{62}{\milli \meter} and \SI{98}{\milli \meter} (compare Fig.~\ref{fig:DSRR-WireFrameComsol}).
The inner length and diameter of the stainless steel (316L) vacuum chamber housing the DSRR are \SI{334}{\milli \meter} and \SI{150}{\milli \meter}, respectively.
We tune our DSRR through exchangeable polymeric gap inserts manufactured to precise and reproducible dimensions by a high-performance hot-lithography 3D printer~\cite{CubicureUrl}.
For the ultimately chosen gap size of \SI{1.12}{\milli \meter} the radio-frequency (RF) response of the DSRR yields a resonance curve with a centroid at \SI{327.3}{\mega \hertz} and a \SI{3}{\decibel} bandwidth of \SI{1.44}{\mega \hertz} equivalent to a quality factor of 227.
Both $\sigma$ transitions are encompassed within the bandwidth of the resonance curve at the value of $B_\text{stat}=\SI{8.4}{\micro \tesla}$, as applied during all measurements reported in the main publication.
Scan ranges of \SI{12}{\kilo \hertz} are much smaller than the bandwidth of the DSRR.

In ideal Rabi spectroscopy the atom's interaction with a $B_\text{osc}$ is only switched on for a well defined time $\tau_\text{int}$.
The RF field of the DSRR comes close to this ideal situation.
Clear deviations only appear at the entrance and exit where the field bends away radially from the central axis as indicated by small arrows in Fig.~\ref{fig:DSRR-WireFrameComsol}.
Small variation in the RF field's strength and $\tau_\text{int}$ result in under or over conversions (deviation from a $\pi$ puse) and alter the line widths.
However, these effects do not lead to asymmetric line-shapes or frequency shifts for the observed transition probabilities. \\

\subsubsection{Static magnetic field generation}
\begin{figure}[h]
    \centering
    \includegraphics[width=0.7\columnwidth]{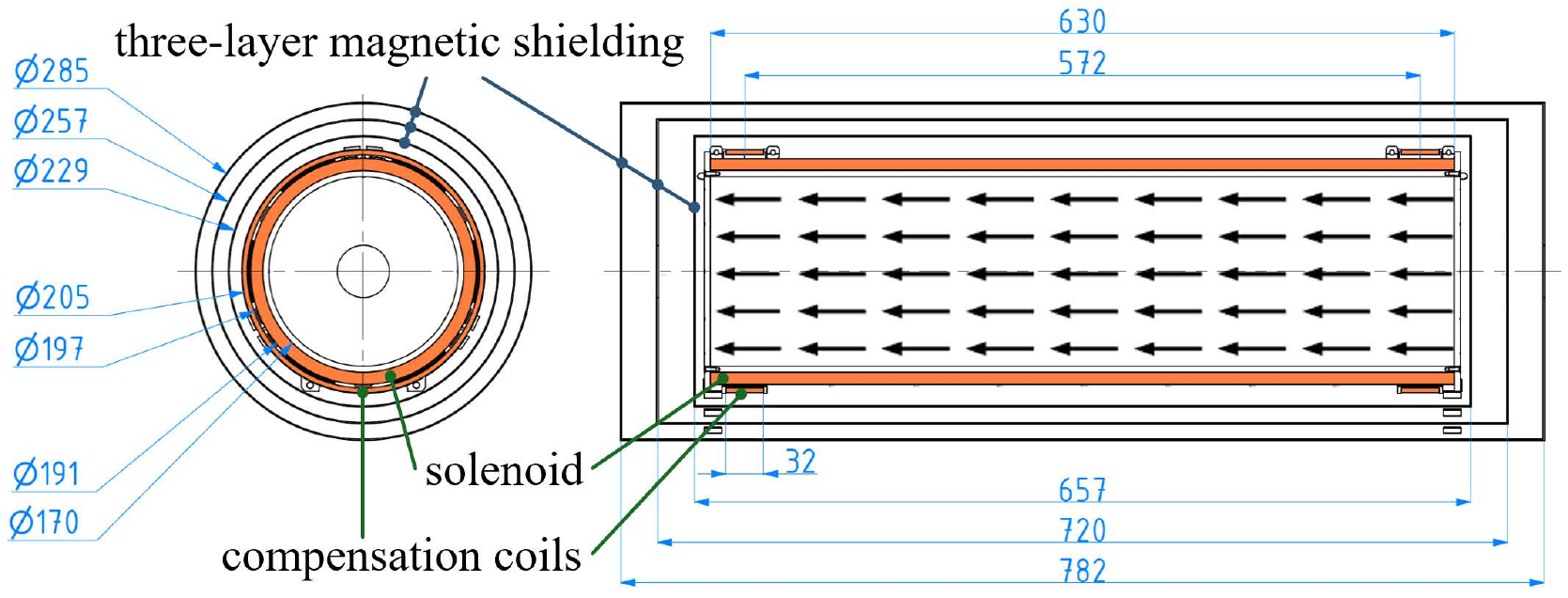}
    \caption{Dimensional cut views of the compensated solenoid and shielding. $B_\text{stat}$ as obtained from Comsol~\cite{ComsolUrl} simulations is depicted by arrows in the longitudinal cut view. Dimensions are given in \SI{}{mm}.}
    \label{fig:Solenoid-WireFrameComsol} 
\end{figure}

Dimensions of the compensated solenoid and the cylindrical three-layer magnetic shielding (non-oriented 80\% nickel-iron-molybdenum alloy) are given in Fig.~\ref{fig:Solenoid-WireFrameComsol}.
Copper wire  of \SI{1.7}{\milli \meter} in diameter was used.
The solenoid has 2068 turns arranged in 6 layers.
A total of 36 compensation turns counteract the field drop at the solenoid's entrance and exit.  
Simulations guiding the design of the solenoid and the three-layer magnetic shielding indicated a field uniformity of $\sigma_B / \overline{B} \sim 10^{-4}$ on axis inside the DSRR.
This is a factor of 5 better, than measured in the constructed solenoid for currents of $\SI{50}{\milli \ampere}$ and $\SI{100}{\milli \ampere}$ ($B_\text{stat}\sim\SI{207}{\micro \tesla}$ and $\SI{414}{\micro \tesla}$).
A high-precision current source with a nominal stability of a few $\sim \! 10^{-6}$ supplies the current in series to the compensated solenoid and a shunt resistor.
The voltage drop across the latter, measured by a precision voltmeter, serves as additional monitor for the generated $B_\text{stat}$.
The hyperfine measurements were performed at smaller current (field) of \SI{2}{\milli \ampere} (\SI{8.4}{\micro \tesla}) than used during the characterisation by field mappings.
In this regime the contribution to inhomogeneities from the background field becomes relevant.
Earth's magnetic field is the dominant external perturbation and shielded with factors of $\sim \! 10^4$ in axial and $\sim \! 10^6$ in radial direction according to Comsol~\cite{ComsolUrl} simulations.
The axial direction is two orders of magnitude worse due to the required axial openings for the beam pipe.
On-axis field measurements of $B_\text{stat}$ with no current applied yield field variations amounting to $\sim$\SI{14}{\nano \tesla}.
This was acceptable for the present measurement, but is also a clear indication for potential future improvements. 
The Earth's magnetic field should be shielded to absolute values below $\sim$\SI{5}{\nano \tesla}, and the associated inhomogeneities should be significantly smaller.
A possible reason for larger variations of the background field could be unidentified residual magnetization of materials of the solenoid and the RF device.
Through the Zeeman slopes the field variations translate to a changing value of the central frequency during the RF interaction.
This effect is the main contributor to the asymmetries observed in the measured resonance spectra (compare Figs.~3~and~5 of main publication). 
The Zeeman slopes are sign-opposite for the two $\sigma$ transitions with a small difference in absolute value due to the $x^2$ terms in Eq.~(1) of the main publication.
At \SI{8.4}{\micro \tesla} the values are \SI{9.357}{\hertz \, \nano \tesla^{-1}} and \SI{-9.321}{\hertz \, \nano \tesla^{-1}} for the $\sigma_1$ and $\sigma_2$ transition, respectively.
Therefore, field variations of $\sim$\SI{14}{\nano \tesla} lead to frequency effects on the order of \SI{140}{\hertz} ($\sim$6 \% of the line width).
The opposite sign of the Zeeman slope causes the line shapes of the two transitions to be mirrored.
At the present level of statistics there is no sensitivity to the subtle relative difference of $\sim$0.4 \% between the absolute values of the two Zeeman slopes.

\subsection{Data acquisition protocol}
The sequence of measurements performed to acquire a resonance pair is as follows:  ref-$\sigma_1^{n_1(1)}$-ref-$\sigma_2^{n_2(1)}$-ref-...-ref-$\sigma_1^{n_1(25)}$-ref-$\sigma_2^{n_2(25)}$-ref, where $n_1(i),n_2(i)$ stand for two randomized sequences.
The reference rates before ($R_\text{ref}^{n-}$) and after ($R_\text{ref}^{n+}$) a transition rate measurement ($R_\sigma^n$) are merged into a time-equivalent average $\overline{R}_\text{ref}^n = \frac{1}{2}(R_\text{ref}^{n-}+R_\text{ref}^{n+})$.
The transition probability at the frequency point $n$ is calculated as $\mathcal{P}^n=1-R_\sigma^n/\overline{R}_\text{ref}^n$.
References are shared ($R_\text{ref}^{n_1(i)+}=R_\text{ref}^{n_2(i)-}$), but not within a single transition spectrum and potential remaining correlations are suppressed by the random sequences.
The acquisition time was \SI{60}{\second} for $R_\sigma$ and \SI{30}{\second} for $R_\text{ref}$.
The total acquisition time per resonance pair including manipulations, reading of set values, and environmental monitoring amounted to \SI{84}{\minute}. \\

\subsection{From sidereal amplitudes to SME constraints}
\subsubsection{Example}
Each amplitude limit provides constraints for the related set of coefficients given by Eq.~(6) of the main publication.
Individual constraints are obtained by allowing only one of the effective coefficients to differ from zero at a time.
As an example we constrain the real part of $H_{p_{211}}^\mathrm{ NR(0B)}$ starting from:
\begin{align}
\mathcal{T}_{p_{211}}^\mathrm{  NR(0B)} =
g_{p_{211}}^\mathrm{  NR(0B)} - H_{p_{211}}^\mathrm{  NR(0B)}=
-\frac{(6 \pi)^{3/2}}{\sin{\vartheta}} \frac{\mathcal{A}^-_1}{\langle|\boldsymbol{p}|^2 \rangle_{(0B)}} .
\tag{S1}
\end{align}
Setting {$\Re(g_{p_{211}}^\mathrm{ NR(0B)})~\equiv~0$} and taking the value and total uncertainty of $\Re(\mathcal{A}^-_{1}) = \SI{-2.5(6.2)}{\hertz}$ as well as $\langle | \boldsymbol{p} |^2 \rangle_{(0B)} = \SI{2.8e-2}{\giga \eV^2}$ (Table~1 of Ref.~\cite{PhysRevD.109.055001}) we obtain \SI{0.6(1.6)e-20}{\giga \eV^{-1}}.
Note that appropriate conversions are required when inserting the experimentally obtained amplitude values in \SI{}{\hertz} into relations to SME coefficients given in natural units.
The steps exemplified here for $\Re ( H_{p_{211}}^\mathrm{ NR(0B)} )$ are listed for all obtained constraints in Table~\ref{tab:SMEcoeffExplicit}.

The angle $\vartheta$ between $B_\text{stat}$ and the Earth's rotation axis can be calculated from the latitude
$\chi' = \frac{\pi}{2} - \chi = 48.7072(2)^\circ$ and the deviation from the north-south alignment
$\phi'_l = \frac{\pi}{2} - \phi_l = 6(2)^\circ$ by using
$\theta_l = -\frac{\pi}{2}$ for Eq.~(24) of Ref.~\cite{PhysRevD.109.055001}:
\begin{align}
\cos{\vartheta} = & \ \cos{\theta_l} \cos{\chi} + \sin{\theta_l} \sin{\chi} \sin{\phi_l} \nonumber \\
= & \ - \cos{\chi'} \cos{\phi'_l}. \nonumber \\
\rightarrow \vartheta = & \ 131.02(18)^\circ .
\tag{S2}
\end{align}
This translates to relative uncertainties on the terms $\sin^{-1}{\vartheta}$, $\sin^{-1}{2\vartheta}$, and $\sin^{-2}{\vartheta}$ of 0.3\%, 0.1\%, and 0.6\%, respectively, which is negligible for the constraints.
Uncertainties on the theoretical values for $\langle | \boldsymbol{p} |^k \rangle_{(qP)}$ are not available, however, there is substantial precedent in the literature on Lorentz violation to disregard those and base the bounds on the experimental uncertainties only. \\

\subsubsection{Detailed list of all SME constraints by the present \deuterium\ hyperfine measurements}
Following the example given above we list all constraints for NR p SME coefficients of the present work and compare to previous constraints in Table~\ref{tab:SMEcoeffExplicit}.
In contrast to the main publication we include the $k=0$ coefficients, where the present \deuterium\ Rabi-type measurements could not improve over existing constraints from \hydrogen\ maser measurements~\cite{Phillips2001}. \\

\begin{table*}
    \caption{Complete list of expressions for constrained SME coefficients with inserted values of amplitudes and p momentum expectation values. The last two columns present the constraint of the present work in comparison to existing results from \hydrogen\ maser measurements.
    }
    \label{tab:SMEcoeffExplicit}
        \renewcommand{\arraystretch}{2.0}
        \setlength{\tabcolsep}{8pt}
            \begin{tabular}{l c r c r c}
                \hline \hline
                $\mathcal{T}^\text{Sun} (\mathcal{A}^{\pm}_m), \mathcal{V}^\text{Sun} (\mathcal{A}^{\pm}_m) \ {}^a$ &
                $k,m,qP$ &
                $C^{\pm}_m, S^{\pm}_m \ {}^b$ &
                $\langle |\boldsymbol{p}|^k \rangle_{(qP)} \ {}^c$ &
                \multicolumn{1}{c}{this work} & \multicolumn{1}{c}{previous$\ {}^d$} \\
                $ \ $ with $\mathcal{A}^{\pm}_m = C^{\pm}_m - i S^{\pm}_m$ &
                & \multicolumn{1}{c}{($\SI{}{\hertz}$)} &
                \multicolumn{1}{c}{($\SI{}{\giga \eV}^{k}$)} &
                \multicolumn{2}{c}{SME constraint in $\SI{}{\giga \eV}^{1-k}$} \\
                \hline 
                $\Re (\mathcal{T}_{p_{011}}^\mathrm{  NR(0B)}) = -\frac{(6 \pi)^{3/2}}{\sin{\vartheta}} \frac{C^-_1}{\langle|\boldsymbol{p}|^0 \rangle_{(0B)}}$ &
                $0,1,0B$ & $2.5 \pm 6.2$ & 1.7 &
                $(-1.0 \pm 2.7) \cdot 10^{-22}$ &  $<9 \cdot 10^{-27}$ \\
                $\Im (\mathcal{T}_{p_{011}}^\mathrm{  NR(0B)}) = +\frac{(6 \pi)^{3/2}}{\sin{\vartheta}} \frac{S^-_1}{\langle|\boldsymbol{p}|^0 \rangle_{(0B)}}$ &
                $0,1,0B$ & $-7.4 \pm 6.1 $ & 1.7 &
                $(-3.1 \pm 2.7) \cdot 10^{-22}$ & $<9 \cdot 10^{-27}$ \\
                $\Re (\mathcal{T}_{p_{011}}^\mathrm{  NR(1B)}) = +\frac{(6 \pi)^{3/2}}{\sin{\vartheta}} \frac{C^-_1}{\langle|\boldsymbol{p}|^0 \rangle_{(1B)}}$ &
                $0,1,1B$ & $2.5 \pm 6.2$ & -3.8 &
                $(-0.5 \pm 1.2) \cdot 10^{-22}$ & $<5 \cdot 10^{-27}$ \\
                $\Im (\mathcal{T}_{p_{011}}^\mathrm{  NR(1B)}) = -\frac{(6 \pi)^{3/2}}{\sin{\vartheta}} \frac{S^-_1}{\langle|\boldsymbol{p}|^0 \rangle_{(1B)}}$ &
                $0,1,1B$ & $-7.4 \pm 6.1 $ & -3.8 &
                $(-1.4 \pm 1.2) \cdot 10^{-22}$ & $<5 \cdot 10^{-27}$ \\
                \hline 
                $\Re (\mathcal{T}_{p_{211}}^\mathrm{  NR(0B)}) = -\frac{(6 \pi)^{3/2}}{\sin{\vartheta}} \frac{C^-_1}{\langle|\boldsymbol{p}|^2 \rangle_{(0B)}}$ &
                $2,1,0B$ & $2.5 \pm 6.2$ & 2.8$\cdot 10^{-2}$ &
                $(-0.6 \pm 1.6) \cdot 10^{-20}$ & $<7 \cdot 10^{-16}$ \\
                $\Im (\mathcal{T}_{p_{211}}^\mathrm{  NR(0B)}) = +\frac{(6 \pi)^{3/2}}{\sin{\vartheta}} \frac{S^-_1}{\langle|\boldsymbol{p}|^2 \rangle_{(0B)}}$ &
                $2,1,0B$ & $-7.4 \pm 6.1 $ & 2.8$\cdot 10^{-2}$ &
                $(-1.9 \pm 1.6) \cdot 10^{-20}$ & $<7 \cdot 10^{-16}$ \\
                $\Re (\mathcal{T}_{p_{211}}^\mathrm{  NR(1B)}) = +\frac{(6 \pi)^{3/2}}{\sin{\vartheta}} \frac{C^-_1}{\langle|\boldsymbol{p}|^2 \rangle_{(1B)}}$ &
                $2,1,1B$ & $2.5 \pm 6.2$ & -1.2$\cdot 10^{-2}$ &
                $(-1.5 \pm 3.7) \cdot 10^{-20}$ & $<4 \cdot 10^{-16}$ \\
                $\Im (\mathcal{T}_{p_{211}}^\mathrm{  NR(1B)}) = -\frac{(6 \pi)^{3/2}}{\sin{\vartheta}} \frac{S^-_1}{\langle|\boldsymbol{p}|^2 \rangle_{(1B)}}$ &
                $2,1,1B$ & $-7.4 \pm 6.1$ & -1.2$\cdot 10^{-2}$ &
                $(-4.4 \pm 3.6) \cdot 10^{-20}$ & $<4 \cdot 10^{-16}$ \\
                \hline 
                $\Re (\mathcal{T}_{p_{411}}^\mathrm{  NR(0B)}) = -\frac{(6 \pi)^{3/2}}{\sin{\vartheta}} \frac{C^-_1}{\langle|\boldsymbol{p}|^4 \rangle_{(0B)}}$ &
                $4,1,0B$ & $2.5 \pm 6.2$ & 9.7$\cdot 10^{-3}$ &
                $(-1.8 \pm 4.6) \cdot 10^{-20}$ & $<9 \cdot 10^{-6}$ \\
                $\Im (\mathcal{T}_{p_{411}}^\mathrm{  NR(0B)}) = +\frac{(6 \pi)^{3/2}}{\sin{\vartheta}} \frac{S^-_1}{\langle|\boldsymbol{p}|^4 \rangle_{(0B)}}$ &
                $4,1,0B$ & $-7.4 \pm 6.1$ & 9.7$\cdot 10^{-3}$ &
                $(-5.4 \pm 4.5) \cdot 10^{-20}$ & $<9 \cdot 10^{-6}$ \\
                $\Re (\mathcal{T}_{p_{411}}^\mathrm{  NR(1B)}) = +\frac{(6 \pi)^{3/2}}{\sin{\vartheta}} \frac{C^-_1}{\langle|\boldsymbol{p}|^4 \rangle_{(1B)}}$ &
                $4,1,1B$ & $2.5 \pm 6.2$ & 3.9$\cdot 10^{-3}$ &
                $(0.5 \pm 1.1) \cdot 10^{-19}$ & $<5 \cdot 10^{-6}$ \\
                $\Im (\mathcal{T}_{p_{411}}^\mathrm{  NR(1B)}) = -\frac{(6 \pi)^{3/2}}{\sin{\vartheta}} \frac{S^-_1}{\langle|\boldsymbol{p}|^4 \rangle_{(1B)}}$ &
                $4,1,1B$ & $-7.4 \pm 6.1 $ & 3.9$\cdot 10^{-3}$ &
                $(1.4 \pm 1.1) \cdot 10^{-19}$ & $<5 \cdot 10^{-6}$ \\
                \hline 
                $\Re (\mathcal{V}_{p_{221}}^\mathrm{  NR}) = -\frac{4 \pi}{\sin{2\vartheta}} \sqrt{\frac{10 \pi}{3}} \frac{C^+_1}{\langle|\boldsymbol{p}|^2 \rangle_{(0E)}}$ &
                $2,1,0E$ & $-4.7 \pm 5.1$ & 7.8$\cdot 10^{-1}$ &
                $(1.6 \pm 1.8) \cdot 10^{-20}$ & --- \\
                $\Im (\mathcal{V}_{p_{221}}^\mathrm{  NR}) = +\frac{4 \pi}{\sin{2\vartheta}} \sqrt{\frac{10 \pi}{3}} \frac{S^+_1}{\langle|\boldsymbol{p}|^2 \rangle_{(0E)}}$ &
                $2,1,0E$ & $-2.7 \pm 5.0$ & 7.8$\cdot 10^{-1}$ &
                $(-0.9 \pm 1.7) \cdot 10^{-20}$ & --- \\
                $\Re (\mathcal{V}_{p_{222}}^\mathrm{  NR}) = -\frac{4 \pi}{\sin{2\vartheta}} \sqrt{\frac{10 \pi}{3}} \frac{C^+_2}{\langle|\boldsymbol{p}|^2 \rangle_{(0E)}}$ &
                $2,2,0E$ & $3.8 \pm 4.9$ & 7.8$\cdot 10^{-1}$ &
                $(-2.3 \pm 3.0) \cdot 10^{-20}$ & --- \\
                $\Im (\mathcal{V}_{p_{222}}^\mathrm{  NR}) = +\frac{4 \pi}{\sin{2\vartheta}} \sqrt{\frac{10 \pi}{3}} \frac{S^+_2}{\langle|\boldsymbol{p}|^2 \rangle_{(0E)}}$ &
                $2,2,0E$ & $3.7 \pm 4.9$ & 7.8$\cdot 10^{-1}$ &
                $(2.2 \pm 3.0) \cdot 10^{-20}$ & --- \\
                \hline 
                $\Re (\mathcal{V}_{p_{421}}^\mathrm{  NR}) = +\frac{4 \pi}{\sin^2{\vartheta}} \sqrt{\frac{10 \pi}{3}} \frac{C^+_1}{\langle|\boldsymbol{p}|^4 \rangle_{(0E)}}$ &
                $4,1,0E$ & $-4.7 \pm 5.1$ & -1.4$\cdot 10^{-3}$ &
                $(-0.9 \pm 1.0) \cdot 10^{-19}$ & --- \\
                $\Im (\mathcal{V}_{p_{421}}^\mathrm{  NR}) = -\frac{4 \pi}{\sin^2{\vartheta}} \sqrt{\frac{10 \pi}{3}} \frac{S^+_1}{\langle|\boldsymbol{p}|^4 \rangle_{(0E)}}$ &
                $4,1,0E$ & $-2.7 \pm 5.0$ & -1.4$\cdot 10^{-3}$ &
                $(0.5 \pm 1.0) \cdot 10^{-19}$ & --- \\
                $\Re (\mathcal{V}_{p_{422}}^\mathrm{  NR}) = -\frac{4 \pi}{\sin^2{\vartheta}} \sqrt{\frac{10 \pi}{3}} \frac{C^+_2}{\langle|\boldsymbol{p}|^4 \rangle_{(0E)}}$ &
                $4,2,0E$ & $3.8 \pm 4.9$ & -1.4$\cdot 10^{-3}$ &
                $(1.3 \pm 1.7) \cdot 10^{-19}$ & --- \\
                $\Im (\mathcal{V}_{p_{422}}^\mathrm{  NR}) = +\frac{4 \pi}{\sin^2{\vartheta}} \sqrt{\frac{10 \pi}{3}} \frac{S^+_2}{\langle|\boldsymbol{p}|^4 \rangle_{(0E)}}$ &
                $4,2,0E$ & $3.7 \pm 4.9 $ & -1.4$\cdot 10^{-3}$ &
                $(-1.2 \pm 1.7) \cdot 10^{-19}$ & --- \\
                \hline \hline
                \multicolumn{6}{l}{$^a$\footnotesize{explicit from Eq.~(6) of main publication} \
                $^b$\footnotesize{from Table~I of main publication,} \
                $^c$\footnotesize{from~\cite{PhysRevD.109.055001},} 
                $^d$\footnotesize{\cite{Phillips2001,PhysRevD.92.056002}} }
            \end{tabular}
\end{table*}

\subsection{Lists of parameters and definitions}
The four subsequent tables list abbreviations (Table~\ref{tab:abbreviations}) as well as definitions, symbols, and constant in context with mathematics and physics (Table~\ref{tab:physmathsymbols}), the hyperfine structure of \deuterium\ (Table~\ref{tab:HFSsymbols}), and the SME framework (Table~\ref{tab:SMEsymbols}). \\

\begin{table*}
    \caption{Alphabetically ordered abbreviations used in the main publication.
    }
    \label{tab:abbreviations}
        \renewcommand{\arraystretch}{1.17}
        \setlength{\tabcolsep}{2pt}
            \begin{tabular}{l l}
                \hline \hline
                Abbreviation & Meaning and Description \\
                \hline
                AD/ELENA & Antiproton Decelerator and Extra Low ENergy Antiproton ring facility of CERN \\
                ASACUSA & collaboration at CERN: Atomic Spectroscopy And Collisions Using Slow Antiprotons \\
                \CPT & combination of the three discrete symmetries: $C$harge conjugation - $P$arity - $T$ime reversal \\
                DSRR & Double-gap Split Ring Resonator \\
                HFS & High-Field-Seekers: atoms in such states experience a force towards higher magnetic fields \\
                LFS & Low-Field-Seekers: atoms in such states experience a force towards lower magnetic fields \\
                REFIMEVE & R\'Eseau FIbré MEtrologique à Vocation Européenne \\  & (metrological fiber network of European vocation) \\
                RF & radio-frequency \\                
                SM & Standard Model of particle physics \\
                SME & Standard-Model Extension: systematic quantitative framework for beyond SM physics \\
                \hline \hline
            \end{tabular}
\end{table*}

\begin{table*}
    \caption{List of physics constants (2022 CODATA values~\cite{Codata2022}) and symbols as well as mathematics notations in order of appearance in main publication.
    }
    \label{tab:physmathsymbols}
        \renewcommand{\arraystretch}{1.17}
        \setlength{\tabcolsep}{2pt}
            \begin{tabular}{c l}
                \hline \hline
                Symbol & Description \\
                \hline
                \hydrogen & hydrogen isotope with proton as nucleus $^1$H\\
                \deuterium & deuterium: hydrogen isotope with a proton and neutron (deuteron) as nucleus $^2$H\\
                $h$ & Planck's constant: $\SI{6.62607015e-34}{\joule \hertz}$ \\
                $g_\text{e}$ & Land\'e $g$ factor of the electron: $\SI{-2.00231930436092(36)}{}$ \\
                $\mu_\text{B}$ & Bohr magneton: $\SI{9.2740100657(29)e-24}{\joule \tesla^{-1}}$ \\
                $g_\text{d}$ & Land\'e $g$ factor of the deuteron: $\SI{0.8574382335(22)}{}$ \\
                $\mu_\text{N}$ & nuclear magneton: $\SI{5.0507837393(16)e-27}{\joule \tesla^{-1}}$ \\
                p, n, e & proton, neutron, electron \\
                $\Re$, $\Im$ & real and imaginary part\\
                $\overline{N}, \varsigma_N$ & denotes mean value ($\overline{\text{overline}}$) and standard deviation (alternative symbol for $\sigma$) of a set of values $N$ \\
                $\chi^2_\text{ndf}$ & reduced $\chi$-square: square-sum of residuals normalized by number of degrees of freedom \\
                \hline \hline
            \end{tabular}
\end{table*}

\begin{table*}
    \caption{Symbols introduced in connection to the hyperfine structure, Rabi-type spectroscopy and the used fit function. Listed in order of appearance in the main publication.
    }
    \label{tab:HFSsymbols}
        \renewcommand{\arraystretch}{1.17}
        \setlength{\tabcolsep}{2pt}
            \begin{tabular}{c l}
                \hline \hline
                Abbrev. or Symbol & Description \\
                \hline
                $F$ & total angular momentum quantum number \\
                $M_F$ & magnetic quantum number \\
                $B_\text{stat}$ & external (static) magnetic field \\
                $\sigma$ transitions & deuterium hyperfine transitions between the $F=\frac{1}{2}$ and $F=\frac{1}{2}$ states with $\Delta M_F = 0$ \\
                $\sigma_1$ transition & deuterium hyperfine transition between states with $M_F=+\frac{1}{2}$ \\
                $\sigma_2$ transition & deuterium hyperfine transition between states with $M_F=-\frac{1}{2}$ \\
                $\nu_{\sigma_1}$, $\nu_{\sigma_2}$ & transition frequency of the $\sigma_1$, $\sigma_2$ transition \\
                $\nu_0^\text{D}$ & zero-field hyperfine structure of deuterium: $\SI{327384352.5222(17)}{\hertz}$ \\
                $x$ & $B_\text{stat}$ in units of a deuterium's characteristic field $B_\text{c}=h \nu_0^D / \mu_-^\text{D} = \SI{11.68}{\milli \tesla}$ \\
                $\mu_-^\text{D}$ & combination of deuteron and electron magnetic moment: $\mu_-^\text{D} = - g_\text{e} \mu_\text{B} - g_\text{d} \mu_\text{N}$ \\
                  & \hspace{0.5cm} appears in the Breit-Rabi equations for deuterium: $\mu_-^\text{D} \sim \SI{1.85652e-23}{\joule \tesla^{-1}} $ \\
                $v$ & velocity \\
                $B_\text{osc}$ & oscillating magnetic field of the RF field: $B_\text{osc} \propto \sqrt{P_\text{RF}}$, where $P_\text{RF}$ is the applied RF power \\
                $\pi$ pulse & drives first population inversion in Rabi oscillations by interaction of matching duration and strength \\
                $R$ & rate of detection of deuterium from the modulated beam \\
                $\mathcal{P}$ & state conversion probability \\
                $\nu^\text{c}$ & observed central frequency of a deuterium hyperfine transition as obtained by a fit \\
                $\nu^\text{c}$, $\lambda$, $\kappa$ & three open fit parameters, $\nu^\text{c}$ defined above, $\lambda$ and $\kappa$ are slope and offset of a linear scaling \\
                $\xi,\Omega_\text{R},\tau_\text{int},\rho,\Omega_\text{L},\Gamma$ & six predetermined parameters (detailed in Table~IV of End Matter in main publication) \\
                $\Delta \Omega$ & detune of variable frequency to central frequency $\Delta\Omega=2\pi(\nu-\nu^\text{c})$ \\
                \hline \hline
            \end{tabular}
\end{table*}

\begin{table*}
    \caption{Notation of the Standard-Model Extension relevant to the present work. Order chosen for sequential readability.  
    }
    \label{tab:SMEsymbols}
    \renewcommand{\arraystretch}{1.45}
    \setlength{\tabcolsep}{2pt}
    \begin{tabular}{c l}
        \hline \hline
        Symbol & Description \\
        \hline
        $\mathcal{T}_{\mathcale{w}_{kjm}}^\text{ NR($qP$)}$ & nonrelativistic (NR) spin-dependent SME coefficients,
        $\mathcal{T}_{\mathcale{w}_{kjm}}^\text{ NR($qP$)} =
    g_{\mathcale{w}_{kjm}}^\text{ NR($qP$)} - H_{\mathcale{w}_{kjm}}^\text{ NR($qP$)} $ \\
        $g_{\mathcale{w}_{kjm}}^\text{ NR($qP$)}$ & \CPTo NR spin-dependent SME coefficients \\
        $H_{\mathcale{w}_{kjm}}^\text{ NR($qP$)}$ & \CPTe NR spin-dependent SME coefficients \\
        $\mathcal{V}_{\mathcale{w}_{kjm}}^\text{ NR}$ & nonrelativistic spin-independent SME coefficients,
        $\mathcal{V}_{\mathcale{w}_{kjm}}^\text{ NR} =
    c_{\mathcale{w}_{kjm}}^\text{ NR}       - a_{\mathcale{w}_{kjm}}^\text{ NR}$\\
        $c_{\mathcale{w}_{kjm}}^\text{ NR}$ & \CPTe NR spin-independent SME coefficients \\
        $a_{\mathcale{w}_{kjm}}^\text{ NR}$ & \CPTo NR spin-independent SME coefficients \\
        NR & nonrelativistic\\
        $\mathcale{w}$ & particle flavour of SME coefficient, which can be p, n, or e in context of deuterium \\
        $k$ & momentum power of SME coefficient \\
        $j$ & spherical tensor rank \\
        $m$ & spherical tensor component; equal to order of siderial variation \\
        $q$ & spin weight \\
        $P$ & parity type, with $(-1)^j$ or $(-1)^{(j-1)}$  indicated by $E$ or $B$, respectively \\
        $\nu_\pm$ & sum ($\nu_+$) and difference ($\nu_-$) of the $\sigma_1$ and $\sigma_2$ transition frequency: $\nu_\pm = \nu_{\sigma_1} \pm \nu_{\sigma_2}$ \\
        $d$ & mass dimension of SME coefficient \\      
        $\boldsymbol{p}$ & proton's relative momentum in nucleus \\
        $\langle |\boldsymbol{p}|^k \rangle_{(qP)}$ & proton momentum expectations values of order $k$ \\
        $\vartheta$ & angle between $B_\text{stat}$ and Earth's rotation axis \\
        $\omega_{\oplus}$ & sidereal frequency \\
        $T_L$ & local sidereal time \\
        $\varphi_\oplus$ & sidereal phase ${\varphi_\oplus = \! \! \!\mod{(\omega_\oplus T_L,2\pi)}}$ \\
        $A_0^\pm$ & constant amplitude of Fourier series fit to sum and difference frequencies $\nu_\pm$ \\
        $C_m^\pm$ & amplitude of order $m$ cosine term of Fourier series fit to sum and difference frequencies $\nu_\pm$ \\
        $S_m^\pm$ & amplitude of order $m$ sine term of Fourier series fit to sum and difference frequencies $\nu_\pm$ \\
        $\mathcal{A}^{\pm}_m$ & definition of a complex amplitude for compact notification: $\mathcal{A}^{\pm}_m = C^{\pm}_m - i S^{\pm}_m$ \\
        \hline \hline
    \end{tabular}
\end{table*}


\begin{thebibliography}{73}
\providecommand{\natexlab}[1]{#1}
\providecommand{\url}[1]{\texttt{#1}}
\expandafter\ifx\csname urlstyle\endcsname\relax
  \providecommand{\doi}[1]{doi: #1}\else
  \providecommand{\doi}{doi: \begingroup \urlstyle{rm}\Url}\fi

\bibitem[Pachucki et~al.(1994)Pachucki, Weitz, and H\"ansch]{Pachucki1994}
K.~Pachucki, M.~Weitz, and T.~W. H\"ansch.
\newblock Theory of the hydrogen-deuterium isotope shift.
\newblock \emph{Phys. Rev. A}, 49:\penalty0 2255--2259, 1994.

\bibitem[Rau et~al.(2020)Rau, Hei{\ss}e, K{\"o}hler-Langes, Sasidharan, Haas, Renisch, D{\"u}llmann, Quint, Sturm, and Blaum]{Rau2020}
S.~Rau, F.~Hei{\ss}e, F.~K{\"o}hler-Langes, S.~Sasidharan, R.~Haas, D.~Renisch, C.~E. D{\"u}llmann, W.~Quint, S.~Sturm, and K.~Blaum.
\newblock Penning trap mass measurements of the deuteron and the hd+ molecular ion.
\newblock \emph{Nature}, 585\penalty0 (7823):\penalty0 43--47, 2020.

\bibitem[Mossa et~al.(2020)Mossa, St{\"o}ckel, Cavanna, Ferraro, Aliotta, Barile, Bemmerer, Best, Boeltzig, Broggini, Bruno, Caciolli, Chillery, Ciani, Corvisiero, Csedreki, Davinson, Depalo, Di~Leva, Elekes, Fiore, Formicola, F{\"u}l{\"o}p, Gervino, Guglielmetti, Gustavino, Gy{\"u}rky, Imbriani, Junker, Kievsky, Kochanek, Lugaro, Marcucci, Mangano, Marigo, Masha, Menegazzo, Pantaleo, Paticchio, Perrino, Piatti, Pisanti, Prati, Schiavulli, Straniero, Sz{\"u}cs, Tak{\'a}cs, Trezzi, Viviani, and Zavatarelli]{Mossa2020}
V.~Mossa, K.~St{\"o}ckel, F.~Cavanna, F.~Ferraro, M.~Aliotta, F.~Barile, D.~Bemmerer, A.~Best, A.~Boeltzig, C.~Broggini, et~al.
\newblock The baryon density of the universe from an improved rate of deuterium burning.
\newblock \emph{Nature}, 587\penalty0 (7833):\penalty0 210--213, 2020.

\bibitem[Nafe and Nelson(1948)]{PhysRev.73.718}
J.~E. Nafe and E.~B. Nelson.
\newblock The hyperfine structure of hydrogen and deuterium.
\newblock \emph{Phys. Rev.}, 73:\penalty0 718--728, 1948.

\bibitem[Pohl et~al.(2010)Pohl, Antognini, Nez, Amaro, Biraben, Cardoso, Covita, Dax, Dhawan, Fernandes, Giesen, Graf, H{\"a}nsch, Indelicato, Julien, Kao, Knowles, Le~Bigot, Liu, Lopes, Ludhova, Monteiro, Mulhauser, Nebel, Rabinowitz, dos Santos, Schaller, Schuhmann, Schwob, Taqqu, Veloso, and Kottmann]{Pohl2010}
R.~Pohl, A.~Antognini, F.~Nez, F.~D. Amaro, F.~Biraben, J.~M.~R. Cardoso, D.~S. Covita, A.~Dax, S.~Dhawan, L.~M.~P. Fernandes, et~al.
\newblock The size of the proton.
\newblock \emph{Nature}, 466\penalty0 (7303):\penalty0 213--216, 2010.

\bibitem[Antognini et~al.(2013)Antognini, Nez, Schuhmann, Amaro, Biraben, Cardoso, Covita, Dax, Dhawan, Diepold, Fernandes, Giesen, Gouvea, Graf, Hänsch, Indelicato, Julien, Kao, Knowles, Kottmann, Bigot, Liu, Lopes, Ludhova, Monteiro, Mulhauser, Nebel, Rabinowitz, dos Santos, Schaller, Schwob, Taqqu, Veloso, Vogelsang, and Pohl]{Antognini2013}
A.~Antognini, F.~Nez, K.~Schuhmann, F.~D. Amaro, F.~Biraben, J.~M.~R. Cardoso, D.~S. Covita, A.~Dax, S.~Dhawan, M.~Diepold, et~al.
\newblock Proton structure from the measurement of 2s-2p transition frequencies of muonic hydrogen.
\newblock \emph{Science}, 339\penalty0 (6118):\penalty0 417--420, 2013.

\bibitem[Pohl et~al.(2016)Pohl, Nez, Fernandes, Amaro, Biraben, Cardoso, Covita, Dax, Dhawan, Diepold, Giesen, Gouvea, Graf, Hänsch, Indelicato, Julien, Knowles, Kottmann, Bigot, Liu, Lopes, Ludhova, Monteiro, Mulhauser, Nebel, Rabinowitz, dos Santos, Schaller, Schuhmann, Schwob, Taqqu, Veloso, Antognini, and Collaboration]{Pohl2016_muD}
R.~Pohl, F.~Nez, L.~M.~P. Fernandes, F.~D. Amaro, F.~Biraben, J.~M.~R. Cardoso, D.~S. Covita, A.~Dax, S.~Dhawan, M.~Diepold, et~al.
\newblock Laser spectroscopy of muonic deuterium.
\newblock \emph{Science}, 353\penalty0 (6300):\penalty0 669--673, 2016.

\bibitem[Gao and Vanderhaeghen(2022)]{Gao2022_muH}
H.~Gao and M.~Vanderhaeghen.
\newblock The proton charge radius.
\newblock \emph{Rev. Mod. Phys.}, 94:\penalty0 015002, 2022.

\bibitem[Hernandez et~al.(2018)Hernandez, Ekström, {Nevo Dinur}, Ji, Bacca, and Barnea]{HERNANDEZ2018377}
O.~Hernandez, A.~Ekström, N.~{Nevo Dinur}, C.~Ji, S.~Bacca, and N.~Barnea.
\newblock The deuteron-radius puzzle is alive: A new analysis of nuclear structure uncertainties.
\newblock \emph{Physics Letters B}, 778:\penalty0 377--383, 2018.

\bibitem[Gotta(2004)]{GOTTA2004133}
D.~Gotta.
\newblock Precision spectroscopy of light exotic atoms.
\newblock \emph{Progress in Particle and Nuclear Physics}, 52\penalty0 (1):\penalty0 133--195, 2004.

\bibitem[Strauch et~al.(2010)Strauch, Amaro, Anagnostopoulos, B\"uhler, Covita, Gorke, Gotta, Gruber, Hirtl, Indelicato, Le~Bigot, Nekipelov, dos Santos, Schlesser, Schmid, Simons, Trassinelli, Veloso, and Zmeskal]{Strauch2010_piD}
T.~Strauch, F.~D. Amaro, D.~F. Anagnostopoulos, P.~B\"uhler, D.~S. Covita, H.~Gorke, D.~Gotta, A.~Gruber, A.~Hirtl, P.~Indelicato, et~al.
\newblock Precision determination of the $d\ensuremath{\pi}\ensuremath{\leftrightarrow}nn$ transition strength at threshold.
\newblock \emph{Phys. Rev. Lett.}, 104:\penalty0 142503, 2010.

\bibitem[Iwasaki et~al.(1997)Iwasaki, Hayano, Ito, Nakamura, Terada, Gill, Lee, Olin, Salomon, Yen, Bartlett, Beer, Mason, Trayling, Outa, Taniguchi, Yamashita, and Seki]{PhysRevLett.78.3067}
M.~Iwasaki, R.~S. Hayano, T.~M. Ito, S.~N. Nakamura, T.~P. Terada, D.~R. Gill, L.~Lee, A.~Olin, M.~Salomon, S.~Yen, et~al.
\newblock Observation of kaonic hydrogen ${K}_{\ensuremath{\alpha}}$ x rays.
\newblock \emph{Phys. Rev. Lett.}, 78:\penalty0 3067--3069, 1997.

\bibitem[Beer et~al.(2005)Beer, Bragadireanu, Cargnelli, Curceanu-Petrascu, Egger, Fuhrmann, Guaraldo, Iliescu, Ishiwatari, Itahashi, Iwasaki, Kienle, Koike, Lauss, Lucherini, Ludhova, Marton, Mulhauser, Ponta, Schaller, Seki, Sirghi, Sirghi, and Zmeskal]{PhysRevLett.94.212302}
G.~Beer, A.~M. Bragadireanu, M.~Cargnelli, C.~Curceanu-Petrascu, J.-P. Egger, H.~Fuhrmann, C.~Guaraldo, M.~Iliescu, T.~Ishiwatari, K.~Itahashi, et~al.
\newblock Measurement of the kaonic hydrogen x-ray spectrum.
\newblock \emph{Phys. Rev. Lett.}, 94:\penalty0 212302, 2005.

\bibitem[Curceanu et~al.(2019)Curceanu, Guaraldo, Iliescu, Cargnelli, Hayano, Marton, Zmeskal, Ishiwatari, Iwasaki, Okada, Sirghi, and Tatsuno]{Curceanu2019KaonicHDReview}
C.~Curceanu, C.~Guaraldo, M.~Iliescu, M.~Cargnelli, R.~Hayano, J.~Marton, J.~Zmeskal, T.~Ishiwatari, M.~Iwasaki, S.~Okada, et~al.
\newblock The modern era of light kaonic atom experiments.
\newblock \emph{Rev. Mod. Phys.}, 91:\penalty0 025006, 2019.

\bibitem[Richard and Sainio(1982)]{RICHARD1982349}
J.~Richard and M.~Sainio.
\newblock Nuclear effects in protonium.
\newblock \emph{Physics Letters B}, 110\penalty0 (5):\penalty0 349--352, 1982.

\bibitem[Wycech et~al.(1985)Wycech, Green, and Niskanen]{WYCECH1985308}
S.~Wycech, A.~Green, and J.~Niskanen.
\newblock On the energy levels in antiprotonic deuterium.
\newblock \emph{Physics Letters B}, 152\penalty0 (5):\penalty0 308--312, 1985.

\bibitem[Baker et~al.(1988)Baker, Batty, Clark, Moir, Sakamoto, Davies, Lowe, Nelson, Pyle, Selvarajah, Squier, Welsh, Winter, Lingeman, {van Eijk}, Hollander, Langerveld, Okx, and Zoutendijk]{BAKER1988631}
C.~Baker, C.~Batty, S.~Clark, J.~Moir, S.~Sakamoto, J.~Davies, J.~Lowe, J.~Nelson, G.~Pyle, A.~Selvarajah, et~al.
\newblock Measurement of x-rays from anti-protonic hydrogen and deuterium.
\newblock \emph{Nuclear Physics A}, 483\penalty0 (3):\penalty0 631--652, 1988.

\bibitem[Augsburger et~al.(1999{\natexlab{a}})Augsburger, Anagnostopoulos, Borchert, Chatellard, Egger, El-Khoury, Gorke, Gotta, Hauser, Indelicato, Kirch, Lenz, Rashid, Siems, and Simons]{AUGSBURGER1999417}
M.~Augsburger, D.~Anagnostopoulos, G.~Borchert, D.~Chatellard, J.-P. Egger, P.~El-Khoury, H.~Gorke, D.~Gotta, P.~Hauser, P.~Indelicato, et~al.
\newblock Measurement of the strong interaction parameters in antiprotonic deuterium.
\newblock \emph{Physics Letters B}, 461\penalty0 (4):\penalty0 417--422, 1999{\natexlab{a}}.

\bibitem[Augsburger et~al.(1999{\natexlab{b}})Augsburger, Anagnostopoulos, Borchert, Chatellard, Egger, El-Khoury, Gorke, Gotta, Hauser, Indelicato, Kirch, Lenz, Rashid, Siems, and Simons]{AUGSBURGER1999149}
M.~Augsburger, D.~Anagnostopoulos, G.~Borchert, D.~Chatellard, J.-P. Egger, P.~El-Khoury, H.~Gorke, D.~Gotta, P.~Hauser, P.~Indelicato, et~al.
\newblock Measurement of the strong interaction parameters in antiprotonic hydrogen and probable evidence for an interference with inner bremsstrahlung.
\newblock \emph{Nuclear Physics A}, 658\penalty0 (2):\penalty0 149--162, 1999{\natexlab{b}}.

\bibitem[Phillips et~al.(2001)Phillips, Humphrey, Mattison, Stoner, Vessot, and Walsworth]{Phillips2001}
D.~F. Phillips, M.~A. Humphrey, E.~M. Mattison, R.~E. Stoner, R.~F.~C. Vessot, and R.~L. Walsworth.
\newblock Limit on lorentz and $\mathrm{CPT}$ violation of the proton using a hydrogen maser.
\newblock \emph{Phys. Rev. D}, 63:\penalty0 111101, 2001.

\bibitem[Humphrey et~al.(2003)Humphrey, Phillips, Mattison, Vessot, Stoner, and Walsworth]{Humphrey2003}
M.~A. Humphrey, D.~F. Phillips, E.~M. Mattison, R.~F.~C. Vessot, R.~E. Stoner, and R.~L. Walsworth.
\newblock Testing cpt and lorentz symmetry with hydrogen masers.
\newblock \emph{Phys. Rev. A}, 68:\penalty0 063807, 2003.

\bibitem[Potvliege et~al.(2023)Potvliege, Nicolson, Jones, and Spannowsky]{PhysRevA.108.052825}
R.~M. Potvliege, A.~Nicolson, M.~P.~A. Jones, and M.~Spannowsky.
\newblock Deuterium spectroscopy for enhanced bounds on physics beyond the standard model.
\newblock \emph{Phys. Rev. A}, 108:\penalty0 052825, 2023.

\bibitem[Kosteleck\'y and Vargas(2015{\natexlab{a}})]{Kostelecky2015}
V.~A. Kosteleck\'y and A.~J. Vargas.
\newblock Lorentz and {CPT} tests with hydrogen, antihydrogen, and related systems.
\newblock \emph{Phys. Rev. D}, 92:\penalty0 056002, 2015{\natexlab{a}}.

\bibitem[Vargas(2024)]{PhysRevD.109.055001}
A.~J. Vargas.
\newblock Prospects for testing $cpt$ and lorentz symmetry with deuterium ground-state zeeman-hyperfine transitions.
\newblock \emph{Phys. Rev. D}, 109:\penalty0 055001, 2024.

\bibitem[Rabi et~al.(1938)Rabi, Zacharias, Millman, and Kusch]{PhysRev.53.318}
I.~I. Rabi, J.~R. Zacharias, S.~Millman, and P.~Kusch.
\newblock A new method of measuring nuclear magnetic moment.
\newblock \emph{Phys. Rev.}, 53:\penalty0 318--318, 1938.

\bibitem[Colladay and Kosteleck\'y(1997)]{PhysRevD.55.6760}
D.~Colladay and V.~A. Kosteleck\'y.
\newblock $\textrm{CPT}$ violation and the {S}tandard {M}odel.
\newblock \emph{Phys. Rev. D}, 55:\penalty0 6760--6774, 1997.

\bibitem[Colladay and Kosteleck\'y(1998)]{PhysRevD.58.116002}
D.~Colladay and V.~A. Kosteleck\'y.
\newblock Lorentz-violating extension of the standard model.
\newblock \emph{Phys. Rev. D}, 58:\penalty0 116002, 1998.

\bibitem[Bluhm et~al.(1999)Bluhm, Kosteleck\'y, and Russell]{PhysRevLett.82.2254}
R.~Bluhm, V.~A. Kosteleck\'y, and N.~Russell.
\newblock $\mathit{CPT}$ and lorentz tests in hydrogen and antihydrogen.
\newblock \emph{Phys. Rev. Lett.}, 82:\penalty0 2254--2257, 1999.

\bibitem[Breit and Rabi(1931)]{PhysRev.38.2082.2}
G.~Breit and I.~I. Rabi.
\newblock Measurement of nuclear spin.
\newblock \emph{Phys. Rev.}, 38:\penalty0 2082--2083, 1931.

\bibitem[Wineland and Ramsey(1972)]{Wineland1972}
D.~J. Wineland and N.~F. Ramsey.
\newblock Atomic deuterium maser.
\newblock \emph{Phys. Rev. A}, 5:\penalty0 821--837, 1972.

\bibitem[Mohr et~al.(2025)Mohr, Newell, Taylor, and Tiesinga]{Codata2022}
P.~J. Mohr, D.~B. Newell, B.~N. Taylor, and E.~Tiesinga.
\newblock Codata recommended values of the fundamental physical constants: 2022.
\newblock \emph{Rev. Mod. Phys.}, 97:\penalty0 025002, 2025.

\bibitem[Kosteleck\'y and Russell(2011)]{RevModPhys.83.11}
V.~A. Kosteleck\'y and N.~Russell.
\newblock Data tables for lorentz and $cpt$ violation.
\newblock \emph{Rev. Mod. Phys.}, 83:\penalty0 11--31, 2011.

\bibitem[SME()]{SME-Tables-arXiv}


\bibitem[Kosteleck\'y and Mewes(2009)]{PhysRevD.80.015020}
V.~A. Kosteleck\'y and M.~Mewes.
\newblock Electrodynamics with lorentz-violating operators of arbitrary dimension.
\newblock \emph{Phys. Rev. D}, 80:\penalty0 015020, 2009.

\bibitem[Kosteleck\'y and Mewes(2013)]{PhysRevD.88.096006}
V.~A. Kosteleck\'y and M.~Mewes.
\newblock Fermions with {L}orentz-violating operators of arbitrary dimension.
\newblock \emph{Phys. Rev. D}, 88:\penalty0 096006, 2013.

\bibitem[Kosteleck\'y and Li(2019)]{PhysRevD.99.056016}
V.~A. Kosteleck\'y and Z.~Li.
\newblock Gauge field theories with lorentz-violating operators of arbitrary dimension.
\newblock \emph{Phys. Rev. D}, 99:\penalty0 056016, 2019.

\bibitem[Kosteleck\'y and Vargas(2018)]{PhysRevD.98.036003}
V.~A. Kosteleck\'y and A.~J. Vargas.
\newblock Lorentz and $cpt$ tests with clock-comparison experiments.
\newblock \emph{Phys. Rev. D}, 98:\penalty0 036003, 2018.

\bibitem[Widmann et~al.(2004)Widmann, Hayano, Hori, and Yamazaki]{Widmann200431}
E.~Widmann, R.~Hayano, M.~Hori, and T.~Yamazaki.
\newblock Measurement of the hyperfine structure of antihydrogen.
\newblock \emph{Nuclear Instruments and Methods in Physics Research Section B: Beam Interactions with Materials and Atoms}, 214:\penalty0 31 -- 34, 2004.
\newblock Low Energy Antiproton Physics (LEAP'03).

\bibitem[Diermaier et~al.(2017)Diermaier, Jepsen, Kolbinger, Malbrunot, Massiczek, Sauerzopf, Simon, Zmeskal, and Widmann]{diermaier2017beam}
M.~Diermaier, C.~Jepsen, B.~Kolbinger, C.~Malbrunot, O.~Massiczek, C.~Sauerzopf, M.~Simon, J.~Zmeskal, and E.~Widmann.
\newblock In-beam measurement of the hydrogen hyperfine splitting and prospects for antihydrogen spectroscopy.
\newblock \emph{Nature communications}, 8\penalty0 (1):\penalty0 1--9, 2017.

\bibitem[Malbrunot et~al.(2019)Malbrunot, Diermaier, Simon, Amsler, {Arguedas Cuendis}, Breuker, Evans, Fleck, Kolbinger, Lanz, Leali, Maeckel, Mascagna, Massiczek, Matsuda, Nagata, Sauerzopf, Venturelli, Widmann, Wiesinger, Yamazaki, and Zmeskal]{MALBRUNOT2019110}
C.~Malbrunot, M.~Diermaier, M.~Simon, C.~Amsler, S.~{Arguedas Cuendis}, H.~Breuker, C.~Evans, M.~Fleck, B.~Kolbinger, A.~Lanz, et~al.
\newblock A hydrogen beam to characterize the {ASACUSA} antihydrogen hyperfine spectrometer.
\newblock \emph{Nuclear Instruments and Methods in Physics Research Section A: Accelerators, Spectrometers, Detectors and Associated Equipment}, 935:\penalty0 110--120, 2019.

\bibitem[Nowak et~al.(2024)Nowak, Malbrunot, Simon, Amsler, {Arguedas Cuendis}, Lahs, Lanz, Nanda, Wiesinger, Wolz, and Widmann]{Nowak2024_PLB-HBeamSME}
L.~Nowak, C.~Malbrunot, M.~Simon, C.~Amsler, S.~{Arguedas Cuendis}, S.~Lahs, A.~Lanz, A.~Nanda, M.~Wiesinger, T.~Wolz, et~al.
\newblock Cpt and lorentz symmetry tests with hydrogen using a novel in-beam hyperfine spectroscopy method applicable to antihydrogen experiments.
\newblock \emph{Physics Letters B}, 858:\penalty0 139012, 2024.

\bibitem[Sup({\natexlab{a}})]{SuppMatA1}
See Supplemental Material section A.1 for details on beam path, which includes Ref.~\cite{OpthosUrl}.

\bibitem[Sup({\natexlab{b}})]{SuppMatA2}
See Supplemental Material sections A.2 for details on oscillating magnetic field, which includes Ref.~\cite{Reynolds1991,CubicureUrl,ComsolUrl}.

\bibitem[Sup({\natexlab{c}})]{SuppMatA3}
See Supplemental Material sections A.3 for details on static magnetic field, which includes Ref.~\cite{ComsolUrl}.

\bibitem[Reynolds et~al.(1991)Reynolds, Hayden, and Hardy]{Reynolds1991}
M.~W. Reynolds, M.~E. Hayden, and W.~N. Hardy.
\newblock Hyperfine resonance of atomic deuterium at 1 k.
\newblock \emph{Journal of Low Temperature Physics}, 84\penalty0 (1):\penalty0 87--108, 1991.

\bibitem[Özşahin et~al.(2017)Özşahin, Şimşek, Ünlü, Kiriş, Köse, Mustaçoğlu, Öztürk, and Akan]{DSRR-1}
G.~Özşahin, T.~Şimşek, M.~Ünlü, O.~Kiriş, S.~Köse, H.~Mustaçoğlu, F.~Öztürk, and V.~Akan.
\newblock Parametric analysis of double-split ring resonator as a reflectarray unit cell.
\newblock In \emph{2017 IEEE International Symposium on Antennas and Propagation \& USNC/URSI National Radio Science Meeting}, pages 101--102, 2017.

\bibitem[Debus and Bolivar(2007)]{DSRR-2}
C.~Debus and P.~H. Bolivar.
\newblock {Frequency selective surfaces for high sensitivity terahertz sensing}.
\newblock \emph{Applied Physics Letters}, 91\penalty0 (18):\penalty0 184102, 2007.

\bibitem[Kaur et~al.(2022)Kaur, Frank, Pinto, Tuckey, and Pottie]{Kaur2022}
N.~Kaur, F.~Frank, J.~Pinto, P.~Tuckey, and P.-E. Pottie.
\newblock A 500-km cascaded white rabbit link for high-performance frequency dissemination.
\newblock \emph{IEEE Transactions on Ultrasonics, Ferroelectrics, and Frequency Control}, 69\penalty0 (2):\penalty0 892--901, 2022.

\bibitem[Cantin et~al.(2021)Cantin, Tønnes, Targat, Amy-Klein, Lopez, and Pottie]{Cantin_2021}
E.~Cantin, M.~Tønnes, R.~L. Targat, A.~Amy-Klein, O.~Lopez, and P.-E. Pottie.
\newblock An accurate and robust metrological network for coherent optical frequency dissemination.
\newblock \emph{New Journal of Physics}, 23\penalty0 (5):\penalty0 053027, 2021.

\bibitem[Ref()]{RefimeveUrl}
https://www.refimeve.fr/index.php/en/.

\bibitem[Sup({\natexlab{d}})]{SuppMatB}
See Supplemental Material section B for details on data acquisition protocol.

\bibitem[Note1()]{Note1}
The theoretical analysis toward the present experiment~\cite {PhysRevD.109.055001} assumes $\vartheta \simeq $\SI {49}{\degree } corresponding to the opposite magnetic field direction than applied in this work.

\bibitem[Sup({\natexlab{e}})]{SuppMatD}
See Supplemental Material section D at [URL will be inserted by publisher] for comprehensive table on SME constraints, which includes Ref.~\cite{PhysRevD.109.055001,Phillips2001,Kostelecky2015}.

\bibitem[Killian et~al.(2023)Killian, Burkley, Blumer, Crivelli, Gustafsson, Hanski, Nanda, Nez, Nesvizhevsky, Reynaud, Schreiner, Simon, Vasiliev, Widmann, and Yzombard]{Killian2023}
C.~Killian, Z.~Burkley, P.~Blumer, P.~Crivelli, F.~P. Gustafsson, O.~Hanski, A.~Nanda, F.~Nez, V.~Nesvizhevsky, S.~Reynaud, et~al.
\newblock Grasian: towards the first demonstration of gravitational quantum states of atoms with a cryogenic hydrogen beam.
\newblock \emph{The European Physical Journal D}, 77\penalty0 (3):\penalty0 50, 2023.

\bibitem[Killian et~al.(2024)Killian, Blumer, Crivelli, Hanski, Kloppenburg, Nez, Nesvizhevsky, Reynaud, Schreiner, Simon, Vasiliev, Widmann, and Yzombard]{killian2024}
C.~Killian, P.~Blumer, P.~Crivelli, O.~Hanski, D.~Kloppenburg, F.~Nez, V.~Nesvizhevsky, S.~Reynaud, K.~Schreiner, M.~Simon, et~al.
\newblock Grasian: shaping and characterization of the cold hydrogen and deuterium beams for the forthcoming first demonstration of gravitational quantum states of atoms.
\newblock \emph{The European Physical Journal D}, 78\penalty0 (10):\penalty0 132, 2024.

\bibitem[Ramsey(1950)]{Ramsey1950}
N.~F. Ramsey.
\newblock A molecular beam resonance method with separated oscillating fields.
\newblock \emph{Phys. Rev.}, 78:\penalty0 695--699, 1950.

\bibitem[Ramsey(1990)]{Ramsey1990experiments}
N.~F. Ramsey.
\newblock Experiments with separated oscillatory fields and hydrogen masers.
\newblock \emph{Reviews of modern physics}, 62\penalty0 (3):\penalty0 541, 1990.

\bibitem[Nanda(2020)]{Amit_CPT2019}
A.~Nanda.
\newblock \emph{Progress Towards Ramsey Hyperfine Spectroscopy in ASACUSA}, pages 201--203.
\newblock worldscientific, 2020.

\bibitem[Opt()]{OpthosUrl}
https://e-opthos.com/microwave-cavities.

\bibitem[Cub()]{CubicureUrl}
https://cubicure.com/en/hot-lithography/.

\bibitem[Com()]{ComsolUrl}
COMSOL Multiphysics$^\text{\textregistered}$, https://www.comsol.com.

\bibitem[Schmidt(1937)]{Schmidt1937}
T.~Schmidt.
\newblock {\"U}ber die magnetischen momente der atomkerne.
\newblock \emph{Zeitschrift f{\"u}r Physik}, 106\penalty0 (5):\penalty0 358--361, 1937.

\bibitem[VanderPlas(2018)]{VanderPlas2018}
J.~T. VanderPlas.
\newblock Understanding the lomb–scargle periodogram.
\newblock \emph{The Astrophysical Journal Supplement Series}, 236\penalty0 (1):\penalty0 16, 2018.

\bibitem[VanderPlas and Željko Ivezic´(2015)]{VanderPlas2015}
J.~T. VanderPlas and Željko Ivezic´.
\newblock Periodograms for multiband astronomical time series.
\newblock \emph{The Astrophysical Journal}, 812\penalty0 (1):\penalty0 18, 2015.

\bibitem[Vanderplas et~al.(2012)Vanderplas, Connolly, Željko Ivezi{\'c}, and Gray]{VanderPlas2012}
J.~Vanderplas, A.~J. Connolly, Željko Ivezi{\'c}, and A.~G. Gray.
\newblock Introduction to astroml: Machine learning for astrophysics.
\newblock \emph{2012 Conference on Intelligent Data Understanding}, pages 47--54, 2012.

\bibitem[Lomb(1976)]{Lomb1976}
N.~R. Lomb.
\newblock Least-squares frequency analysis of unequally spaced data.
\newblock \emph{Astrophysics and Space Science}, 39\penalty0 (2):\penalty0 447--462, 1976.

\bibitem[{Scargle}(1982)]{Scargle1982}
J.~D. {Scargle}.
\newblock {Studies in astronomical time series analysis. II. Statistical aspects of spectral analysis of unevenly spaced data.}
\newblock \emph{\apj}, 263:\penalty0 835--853, 1982.

\bibitem[{Astropy Collaboration} et~al.(2022){Astropy Collaboration}, {Price-Whelan}, {Lim}, {Earl}, {Starkman}, {Bradley}, {Shupe}, {Patil}, {Corrales}, {Brasseur}, {N{"o}the}, {Donath}, {Tollerud}, {Morris}, {Ginsburg}, {Vaher}, {Weaver}, {Tocknell}, {Jamieson}, {van Kerkwijk}, {Robitaille}, {Merry}, {Bachetti}, {G{"u}nther}, {Aldcroft}, {Alvarado-Montes}, {Archibald}, {B{'o}di}, {Bapat}, {Barentsen}, {Baz{'a}n}, {Biswas}, {Boquien}, {Burke}, {Cara}, {Cara}, {Conroy}, {Conseil}, {Craig}, {Cross}, {Cruz}, {D'Eugenio}, {Dencheva}, {Devillepoix}, {Dietrich}, {Eigenbrot}, {Erben}, {Ferreira}, {Foreman-Mackey}, {Fox}, {Freij}, {Garg}, {Geda}, {Glattly}, {Gondhalekar}, {Gordon}, {Grant}, {Greenfield}, {Groener}, {Guest}, {Gurovich}, {Handberg}, {Hart}, {Hatfield-Dodds}, {Homeier}, {Hosseinzadeh}, {Jenness}, {Jones}, {Joseph}, {Kalmbach}, {Karamehmetoglu}, {Ka{l}uszy{'n}ski}, {Kelley}, {Kern}, {Kerzendorf}, {Koch}, {Kulumani}, {Lee}, {Ly}, {Ma}, {MacBride}, {Maljaars}, {Muna}, {Murphy}, {Norman}, {O'Steen},
  {Oman}, {Pacifici}, {Pascual}, {Pascual-Granado}, {Patil}, {Perren}, {Pickering}, {Rastogi}, {Roulston}, {Ryan}, {Rykoff}, {Sabater}, {Sakurikar}, {Salgado}, {Sanghi}, {Saunders}, {Savchenko}, {Schwardt}, {Seifert-Eckert}, {Shih}, {Jain}, {Shukla}, {Sick}, {Simpson}, {Singanamalla}, {Singer}, {Singhal}, {Sinha}, {Sip{H{o}}cz}, {Spitler}, {Stansby}, {Streicher}, {{
{S}}umak}, {Swinbank}, {Taranu}, {Tewary}, {Tremblay}, {Val-Borro}, {Van Kooten}, {Vasovi{'c}}, {Verma}, {de Miranda Cardoso}, {Williams}, {Wilson}, {Winkel}, {Wood-Vasey}, {Xue}, {Yoachim}, {Zhang}, {Zonca}, and {Astropy Project Contributors}]{astropy:2022}
{Astropy Collaboration}, A.~M. {Price-Whelan}, P.~L. {Lim}, N.~{Earl}, N.~{Starkman}, L.~{Bradley}, D.~L. {Shupe}, A.~A. {Patil}, L.~{Corrales}, C.~E. {Brasseur}, et~al.
\newblock {The Astropy Project: Sustaining and Growing a Community-oriented Open-source Project and the Latest Major Release (v5.0) of the Core Package}.
\newblock \emph{\apj}, 935\penalty0 (2):\penalty0 167, 2022.

\bibitem[{Astropy Collaboration} et~al.(2018){Astropy Collaboration}, {Price-Whelan}, {Sip{\H{o}}cz}, {G{\"u}nther}, {Lim}, {Crawford}, {Conseil}, {Shupe}, {Craig}, {Dencheva}, {Ginsburg}, {Vand erPlas}, {Bradley}, {P{\'e}rez-Su{\'a}rez}, {de Val-Borro}, {Aldcroft}, {Cruz}, {Robitaille}, {Tollerud}, {Ardelean}, {Babej}, {Bach}, {Bachetti}, {Bakanov}, {Bamford}, {Barentsen}, {Barmby}, {Baumbach}, {Berry}, {Biscani}, {Boquien}, {Bostroem}, {Bouma}, {Brammer}, {Bray}, {Breytenbach}, {Buddelmeijer}, {Burke}, {Calderone}, {Cano Rodr{\'\i}guez}, {Cara}, {Cardoso}, {Cheedella}, {Copin}, {Corrales}, {Crichton}, {D'Avella}, {Deil}, {Depagne}, {Dietrich}, {Donath}, {Droettboom}, {Earl}, {Erben}, {Fabbro}, {Ferreira}, {Finethy}, {Fox}, {Garrison}, {Gibbons}, {Goldstein}, {Gommers}, {Greco}, {Greenfield}, {Groener}, {Grollier}, {Hagen}, {Hirst}, {Homeier}, {Horton}, {Hosseinzadeh}, {Hu}, {Hunkeler}, {Ivezi{\'c}}, {Jain}, {Jenness}, {Kanarek}, {Kendrew}, {Kern}, {Kerzendorf}, {Khvalko}, {King}, {Kirkby}, {Kulkarni},
  {Kumar}, {Lee}, {Lenz}, {Littlefair}, {Ma}, {Macleod}, {Mastropietro}, {McCully}, {Montagnac}, {Morris}, {Mueller}, {Mumford}, {Muna}, {Murphy}, {Nelson}, {Nguyen}, {Ninan}, {N{\"o}the}, {Ogaz}, {Oh}, {Parejko}, {Parley}, {Pascual}, {Patil}, {Patil}, {Plunkett}, {Prochaska}, {Rastogi}, {Reddy Janga}, {Sabater}, {Sakurikar}, {Seifert}, {Sherbert}, {Sherwood-Taylor}, {Shih}, {Sick}, {Silbiger}, {Singanamalla}, {Singer}, {Sladen}, {Sooley}, {Sornarajah}, {Streicher}, {Teuben}, {Thomas}, {Tremblay}, {Turner}, {Terr{\'o}n}, {van Kerkwijk}, {de la Vega}, {Watkins}, {Weaver}, {Whitmore}, {Woillez}, {Zabalza}, and {Astropy Contributors}]{astropy:2018}
{Astropy Collaboration}, A.~M. {Price-Whelan}, B.~M. {Sip{\H{o}}cz}, H.~M. {G{\"u}nther}, P.~L. {Lim}, S.~M. {Crawford}, S.~{Conseil}, D.~L. {Shupe}, M.~W. {Craig}, N.~{Dencheva}, et~al.
\newblock {The Astropy Project: Building an Open-science Project and Status of the v2.0 Core Package}.
\newblock \emph{The Astronomical Journal}, 156\penalty0 (3):\penalty0 123, 2018.

\bibitem[{Astropy Collaboration} et~al.(2013){Astropy Collaboration}, {Robitaille}, {Tollerud}, {Greenfield}, {Droettboom}, {Bray}, {Aldcroft}, {Davis}, {Ginsburg}, {Price-Whelan}, {Kerzendorf}, {Conley}, {Crighton}, {Barbary}, {Muna}, {Ferguson}, {Grollier}, {Parikh}, {Nair}, {Unther}, {Deil}, {Woillez}, {Conseil}, {Kramer}, {Turner}, {Singer}, {Fox}, {Weaver}, {Zabalza}, {Edwards}, {Azalee Bostroem}, {Burke}, {Casey}, {Crawford}, {Dencheva}, {Ely}, {Jenness}, {Labrie}, {Lim}, {Pierfederici}, {Pontzen}, {Ptak}, {Refsdal}, {Servillat}, and {Streicher}]{astropy:2013}
{Astropy Collaboration}, T.~P. {Robitaille}, E.~J. {Tollerud}, P.~{Greenfield}, M.~{Droettboom}, E.~{Bray}, T.~{Aldcroft}, M.~{Davis}, A.~{Ginsburg}, A.~M. {Price-Whelan}, et~al.
\newblock {Astropy: A community Python package for astronomy}.
\newblock \emph{Astron. Astrophys.}, 558, A33, 2013.

\bibitem[Dreissen et~al.(2022)Dreissen, Yeh, F{\"u}rst, Grensemann, and Mehlst{\"a}ubler]{Dreissen2022}
L.~S. Dreissen, C.-H. Yeh, H.~A. F{\"u}rst, K.~C. Grensemann, and T.~E. Mehlst{\"a}ubler.
\newblock Improved bounds on lorentz violation from composite pulse ramsey spectroscopy in a trapped ion.
\newblock \emph{Nature Communications}, 13\penalty0 (1):\penalty0 7314, 2022.

\bibitem[Baluev(2008)]{BaluevMethod}
R.~V. Baluev.
\newblock Assessing the statistical significance of periodogram peaks.
\newblock \emph{Monthly Notices of the Royal Astronomical Society}, 385\penalty0 (3):\penalty0 1279--1285, 2008.

\bibitem[Kosteleck\'y and Vargas(2015{\natexlab{b}})]{PhysRevD.92.056002}
V.~A. Kosteleck\'y and A.~J. Vargas.
\newblock Lorentz and {CPT} tests with hydrogen, antihydrogen, and related systems.
\newblock \emph{Phys. Rev. D}, 92:\penalty0 056002, 2015{\natexlab{b}}.

\end{thebibliography}
\end{document}